\documentclass[aps,prd,english,preprintnumbers,nofootinbib,floatfix,twocolumn,10pt]{revtex4}

\usepackage{amsfonts,amsmath,amssymb}
\usepackage{graphicx,epsfig}
\usepackage{float}
\usepackage{subfig}
\usepackage{subfloat}
\usepackage[utf8]{inputenc}
\usepackage{hyperref}
\usepackage{babel}
\usepackage{array}

\def	\beq	{\begin{equation}}
\def	\eeq	{\end{equation}}
\def	\be	{\begin{equation}}
\def	\ee	{\end{equation}}

\def \tt {\tilde{t}}

\begin{document}

\title{$D\to4$ Einstein-Gauss-Bonnet Gravity and Beyond}

\author{Damien A.~Easson}
\email{deasson@asu.edu }

\author{Tucker Manton}
\email{tucker.manton@asu.edu}

\author{Andrew Svesko}
\email{andrew.svesko@asu.edu}

\affiliation{Department of Physics, Arizona State University \\ Tempe, Arizona 85287, USA\\}

\begin{abstract}
\begin{center}
{\bf Abstract}
\end{center}
\noindent
A `novel' pure theory of Einstein-Gauss-Bonnet gravity in four-spacetime dimensions can be constructed by rescaling the Gauss-Bonnet coupling constant, seemingly eluding Lovelock's theorem. Recently, however, the well-posedness of this model has been called into question. Here we apply a `dimensional regularization' technique, first used by Mann and Ross to write down a $D\to2$ limit of general relativity, to the case of pure Einstein-Gauss-Bonnet gravity. The resulting four-dimensional action is a particular Horndeski theory of gravity matching the result found via a Kaluza-Klein reduction over a flat internal space. Some cosmological solutions of this four-dimensional theory are examined. We further adapt the technique to higher curvature Lovelock theories of gravity, as well as a low-energy effective string action with an $\alpha'$ correction. With respect to the $D\to4$ limit of the $\alpha'$-corrected string action, we find we must also rescale the dilaton to have a non-singular action in four dimensions.  Interestingly, when the conformal rescaling $\Phi$ is interpreted as another dilaton, the regularized string action appears to be a special case of a covariant multi-Galileon theory of gravity.
 \end{abstract}

\thispagestyle{empty}


\maketitle


\section{Introduction} \label{sec:intro}
\indent

Einstein's theory of general relativity (GR) has been tremendously successful at describing astrophysical phenomena, as well as the evolution of the universe. What's more, it is the unique covariant theory of gravity in $D=4$ spacetime dimensions with second order equations of motion of the metric without any ghost-like instabilities. The feature of uniqueness is a consequence of  Lovelock's theorem \cite{Lovelock:1971yv,Lovelock:1972vz}. This states that Lovelock theories of gravity are the unique, pure covariant theories of gravity with second order equations of motion whose (Euclideanized) action $I_{(2n)}$ is the Euler-characteristic $\chi(\mathcal{M}_{2n})$ of a compact $2n$-dimensional  manifold $\mathcal{M}_{2n}$, such that in $D=2n$ it is purely topological, and therefore does not contribute to local spacetime dynamics. In $D=3,4$, GR is the only Lovelock theory which is non-trivial; in $D=2$ the Einstein-Hilbert Lagrangian is the Euler-density for a two-dimensional manifold and is thus trivial -- indeed, in $D=2$ the Einstein tensor vanishes identically, $G_{\mu\nu}=0$. 

In dimensions $D\geq5$,  the Lovelock Lagrangians include higher curvature corrections to GR and yield interesting physics. The simplest such correction is the Gauss-Bonnet contribution, leading to Einstein-Gauss-Bonnet gravity (EGB), with action,
\beq
\begin{split}
 I&=\int d^{D}x\sqrt{-g}\biggr[\frac{1}{16\pi G_{D}}(R-2\Lambda_{0})+\alpha_{GB}\mathcal{L}_{GB}\biggr]\label{IEGBintro}
\end{split}
\eeq
where $\alpha_{GB}$ is the Gauss-Bonnet coupling, $\mathcal{L}_{GB}=(R^{2}-4R_{\mu\nu}^{2}+R_{\mu\nu\rho\sigma}^{2})$ is the Gauss-Bonnet curvature, and $\Lambda_{0}$ is the cosmological constant. The equations of motion of the theory are, when we include matter, 
\beq \frac{1}{8\pi G_{D}}(G_{\mu\nu}+\Lambda_{0}g_{\mu\nu})+\alpha_{GB}\mathcal{H}_{\mu\nu}=T_{\mu\nu}\;,\eeq
where
\beq 
\begin{split}
\mathcal{H}_{\mu\nu}&=2\biggr[RR_{\mu\nu}-2R_{\mu\alpha\nu\beta}R^{\alpha\beta}+R_{\mu\alpha\beta\sigma}R_{\nu}^{\;\alpha\beta\sigma}\\
&-2R_{\mu\alpha}R^{\alpha}_{\;\nu}-\frac{1}{4}g_{\mu\nu}\mathcal{L}_{GB}\biggr]\;.\label{partofEGBeom}
\end{split}
\eeq
In $D=4$ the Gauss-Bonnet contribution is purely topological and does not alter the local dynamics of the theory\footnote{This is not to say that the Gauss-Bonnet does not have \emph{any} effect in four dimensions. Indeed, a Gauss-Bonnet term alters the form of the horizon entropy  by a constant proportional to the 2D Euler character of the horizon, and thus naively leads to a violation of the second law of thermodynamics during black hole mergers \cite{Chatterjee:2013daa}.}.  This can be verified by studying the trace of the equations of motion
\beq \frac{1}{8\pi G_{D}}\left[\left(1-\frac{D}{2}\right)R+D\Lambda_{0}\right]+2\alpha_{GB}\left(1-\frac{D}{4}\right)\mathcal{L}_{GB}=T\;.\label{traceEGB1}\eeq
In $D=4$ the Gauss-Bonnet contribution vanishes.

Recently \cite{Glavan:2019inb} uncovered a $D=4$ limit of Einstein-Gauss-Bonnet gravity. This so-called `novel EGB' model is derived by a simple rescaling of the Gauss-Bonnet coupling $\alpha_{GB}$ by a term proportional to the dimension in which it becomes purely topological, \emph{i.e.},
\beq \alpha_{GB}\to (D-4)\alpha_{GB}\;.\label{GBcouplingrescale} \eeq
This dimensional rescaling trick can be easily adapted to Lovelock theories more broadly \cite{Casalino:2020kbt}, as we will describe in more detail later.

The resulting theory is a pure covariant theory of gravity that has second order equations of motion for which the Gauss-Bonnet contribution affects the local dynamics, yielding a non-trivial modification to general relativity in four dimensions. In recent months, novel EGB has been explored, including studies on black hole physics \cite{Glavan:2019inb,Fernandes:2020rpa,Konoplya:2020qqh,Kumar:2020owy,Ghosh:2020vpc,Ghosh:2020syx,Kumar:2020uyz,Kumar:2020xvu,Heydari-Fard:2020sib}, their (extended) thermodynamics \cite{Hegde:2020xlv,HosseiniMansoori:2020yfj,Wei:2020poh,EslamPanah:2020hoj} and quasinormal modes \cite{Konoplya:2020bxa,Churilova:2020aca,Mishra:2020gce,Aragon:2020qdc}; stars \cite{Glavan:2019inb,Doneva:2020ped,Banerjee:2020stc}; cosmology \cite{Glavan:2019inb,1795133,Odintsov:2020sqy,Odintsov:2020zkl}, and wormholes \cite{Jusufi:2020yus,Liu:2020yhu}. A Hamltonian analysis was also given recently in \cite{Aoki:2020lig}.

Novel EGB, in the form derived by \cite{Glavan:2019inb}, is not without its criticism, however. First off, if we perform the rescaling (\ref{GBcouplingrescale}) and take the $D\to4$ limit in the action (\ref{IEGBintro}), we find that the action itself is singular, such that it is not well-defined locally (this issue also does not disappear if we take take the limit only after we have varied the action \cite{Arrechea:2020evj}). Consequently, without a well-defined local action it is unclear how to count the dynamical degrees of freedom at the non-linear level using a Hamiltonian analysis. Indeed, \cite{Aoki:2020lig} show that the $D\to4$ limit is subtle and depends on the regularization scheme of the Hamiltonian and equations of motion, such that in order to find a consistent theory in four dimensions with the proper two dynamical degrees of freedom, the temporal diffeomorphism invariance of the theory must be broken. Secondly, as pointed out in \cite{Lu:2020iav}, the horizon entropy computed using the Wald entropy functional diverges in the $D\to4$ limit with the rescaled Gauss-Bonnet coupling. Moreover, it is unclear whether the novel EGB theory has  an intrinsic four-dimensional description in terms of a covariantly rank-2 tensor \cite{Gurses:2020ofy}, and also suffers from  an ``index problem'' \cite{Ai:2020peo}.  Additional issues with the novel EGB model have been raised concerning the on-shell action, and the computation of counterterms \cite{Mahapatra:2020rds}. The overall conclusion is that a pure four-dimensional EGB theory is untenable. 

 Collectively, these issues led \cite{Lu:2020iav} to propose an alternative way to take the $D=4$ limit of EGB, accomplished via a Kaluza-Klein-like dimensional reduction over a maximally symmetric internal space \cite{VanAcoleyen11-1,Charmousis:2014mia}  in addition to the rescaling (\ref{GBcouplingrescale}). A particularly striking feature of this alternative $D=4$ EGB theory is that the resulting action is not a pure theory of gravity, but rather is a special case of Horndeski gravity \cite{Horndeski:1974wa}, where the scalar field arises as a component of the $D$-dimensional metric.

In this note we follow a different road to a $D=4$ limit of EGB, as well as its higher dimensional Lovelock counterparts. Our approach\footnote{As this article was in preparation we became aware of the recent work by \cite{Fernandes:2020nbq,Hennigar:2020lsl}, for which we share overlapping ideas.} follows the spirit of Mann and Ross \cite{Mann:1992ar}, who uncovered a non-trivial $D=2$ limit of general relativity by performing a type of ``dimensional regularization," involving subtracting a conformally transformed Einstein-Hilbert term in addition to a rescaling of the $D$-dimensonal Newton's constant $G_{D}\to (1-\frac{D}{2})G_{D},$ reminiscent of (\ref{GBcouplingrescale}). As we will review later, the resulting gravitational action is described by an Einstein-dilaton theory of gravity, matching a dimensional reduction of $D=4$ GR, and whose equations of motion yield an $R=T$ theory, where $R$ and $T$  are the 2-dimensional Ricci scalar and energy-momentum tensor, respectively. Likewise, we will find that our dimensionally regularized EGB model results in a Horndeski theory of gravity that is intimately related to the dimensionally reduced model of \cite{Lu:2020iav} (equivalent, in fact, after some simple field redefinitions and for a flat internal space), and has the proper trace equation. 

Aside from Lovelock's theorem, there are other fundamental reasons why the Gauss-Bonnet contribution is of interest. In particular, it naturally appears as the $\alpha'$ correction to the gravi-dilaton sector of the low-energy effective action of $10$-dimensional heterotic string theory \cite{Zwiebach:1985uq,Duff86-1,Metsaev87-1}, in order to rid the low-energy theory of any ghosts. It is therefore natural to wonder whether we may write down a $D\to4$ limit of the low-energy effective string action \emph{without} performing a dimensional reduction. As we will show, when the 10-dimensional dilaton is dynamical, we must perform an additional field redefinition in order to have a well-defined $D\to4$ limit. 

The layout of this article is as follows. We set the stage in Section \ref{sec:D2limGR} with a very brief review of the $D\to2$ limit of GR worked out by Mann and Ross in \cite{Mann:1992ar}, where we also show how to rewrite the resulting Einstein-dilaton theory as a pure theory of gravity. In Section \ref{sec:D4limEGB} we adapt the Mann-Ross `dimensional regularization' technique and write down a $D\to4$ limit of Einstein-Gauss-Bonnet gravity, which we also relate to the $D=4$ EGB model uncovered via Kaluza-Klein reduction \cite{Lu:2020iav}. This section has strong overlap with the recent work \cite{Fernandes:2020nbq,Hennigar:2020lsl}. We show how to apply the regularization technique to more general Lovelock theories of gravity in Section \ref{sec:novelLL} and the low energy effective action of a string theory in Section \ref{sec:effstring}. We find in the $D\to4$ limit of the string action that we must also rescale the dilaton to avoid any singularities in the action. Section \ref{sec:cosmo} is devoted to a brief analysis of cosmological solutions of the resulting $D=4$ Horndeski theory of gravity uncovered in Section \ref{sec:D4limEGB} and  \cite{Lu:2020iav}, where we find a class of constant curvature solutions for $\Phi$ being linear in cosmological time, as well as study the late time behavior. We conclude in Section \ref{sec:disc} with a discussion on multiple avenues for future work.


\section{The $D\to2$ Limit of General Relativity} \label{sec:D2limGR}
\indent

Consider Einstein gravity in $D$-spacetime dimensions:
\beq I_{EH}=\int d^{D}x\sqrt{-g}\left[\frac{1}{16\pi G_{D}}(R-2\Lambda_{0})+\mathcal{L}^{(D)}_{\text{mat}}\right]\;.\eeq
The equations of motion are
\beq \frac{1}{8\pi G_{D}}\left(G_{\mu\nu}+\Lambda_{0}g_{\mu\nu}\right)=T_{\mu\nu}\;,\label{GReomD}\eeq
where $T_{\mu\nu}\equiv-\frac{2}{\sqrt{-g}}\frac{\delta S_{mat}}{\delta g^{\mu\nu}}$ is the energy-momentum tensor associated with the matter action $S_{mat}=\int d^{D}x\sqrt{-g}\mathcal{L}_{mat}$. Taking the trace of (\ref{GReomD}) we find that Einstein gravity is trivial in $D=2$ spacetime dimensions:
\beq R\left(1-\frac{D}{2}\right)+D\Lambda_{0}=T \to \Lambda_{0}=8\pi G_{D} T\;.\label{traceGR1}\eeq
This is not so surprising, as in 2-dimensions the Einstein-Hilbert action is the Euler-characteristic for a $2$-dimensional compact spacetime.

Notice, however, what happens when we rescale $G_{D}\to G_{D}\left(1-\frac{D}{2}\right)$. Then, we have a non-trivial $D\to2$ limit of Einstein gravity such that the trace of the field equations (\ref{traceGR1}) exhibits local dynamics
\beq R+D\Lambda_{0}=8\pi G_{D}T \to R+2\Lambda_{0}=8\pi G_{2}T\;,\eeq
where $G_{2}$ is the 2-dimensional Newton's constant. 

This issue can seemingly be resolved following a method developed by Mann and Ross \cite{Mann:1992ar}, where they consider the Einstein-Hilbert action in $D$-dimensions, subtract off a conformally related Einstein-Hilbert action, and then perform a power series expansion in $(D-2)$, eventually taking the limit $D\to2$, together with a rescaling of $G_{D}$ as above. More explicitly, subtract a term $\int d^{D}x\sqrt{-\tilde{g}}\tilde{R}/16\pi G_{D}$ such that it becomes a total derivative in $D=2$, i.e., $\lim_{D\to2}\tilde{G}_{\mu\nu}/16\pi G_{D}=0$. Moreover, we assume the metric $\tilde{g}_{\mu\nu}$ is conformally related to $g_{\mu\nu}$ via $\tilde{g}_{\mu\nu}=e^{\Phi}g_{\mu\nu}$, such that 
\beq 
\begin{split}
\tilde{R}&=e^{-\Phi}\biggr(R-(D-1)g^{\mu\nu}\nabla_{\mu}\nabla_{\nu}\Phi\\
&-\frac{1}{4}(D-2)(D-1)g^{\mu\nu}\nabla_{\mu}\Phi\nabla_{\nu}\Phi\biggr)\;,
\end{split}
\eeq
and $\sqrt{-\tilde{g}}=e^{D\Phi/2}\sqrt{-g}$.  Then consider the action
\beq I=\frac{1}{16\pi G_{D}}\int d^{D}x\sqrt{-g}\left(R-2\Lambda_{0}-\frac{\sqrt{-\tilde{g}}}{\sqrt{-g}}\tilde{R}\right)+I_{\text{mat}}\;.\eeq
At this point this action is equivalent to the $2$-dimensional Einstein-Hilbert action in the limit $D\to2$, by construction. Using the conformal rescaling, performing a power series expansion in $e^{(D/2-1)\Phi}$ to order $O(D/2-1)$ dropping any total derivatives, rescaling Newton's constant $\lim_{D\to2}G_{D}\to\left(1-\frac{D}{2}\right)G_{2}$, and finally taking the $D\to2$ limit, we have
\beq I=\frac{1}{16\pi G_{2}}\int d^{2}x\sqrt{-g}(\Phi R -2\Lambda_{0}+\frac{1}{2}(\nabla\Phi)^{2})+I_{\text{mat}}\;.\label{Mannact2}\eeq

The equations of motion for this action are
\beq
\begin{split}
 0&=\frac{1}{2}\left(\nabla_{\mu}\Phi\nabla_{\nu}\Phi-\frac{1}{2}g_{\mu\nu}(\nabla\Phi)^{2}\right)-\nabla_{\mu}\nabla_{\nu}\Phi  +g_{\mu\nu}\nabla^{2}\Phi \\
 &+g_{\mu\nu}\Lambda_{0}-8\pi G_{2}T_{\mu\nu}\;,\label{D2GReom1}
\end{split}
\eeq
\beq 0=R-\nabla^{2}\Phi\;.\label{D2GReom2}\eeq
The second of these provides a constraint such that the trace of the first leads to an $R=T$ theory (plus cosmological constant):
\beq R=8\pi G_{2}T-2\Lambda_{0}\;.\label{D2GReomv1}\eeq

While the action (\ref{Mannact2}) is a 2D gravi-dilaton model, it is special in that its equations of motion lead to what can be interpreted as the $D\to2$ limit of Einstein's field equations (\ref{D2GReomv1}), and so (\ref{Mannact2}) can reasonably be considered the closest thing to general relativity that exists in two dimensions. 
Moreover, \cite{Mann:1992ar} showed that (\ref{Mannact2}) also arises from a typical dimension reduction (plus rescaling of coupling) of the Einstein-Hilbert action in $D=4$ spacetime dimensions, much like its stringy motivated counterparts (it is in this sense that $\Phi$ should be interpreted as a dilaton). The approach of \cite{Mann:1992ar} has since been generalized to derive 2D Liouville gravity from this type of rescaling \cite{Grumiller:2007wb}. The  model described by (\ref{Mannact2}) was also recently used to provide a $D\to2$ limit of extended black hole thermodynamics \cite{Frassino:2015oca}. 

As noted in \cite{Rosso:2020zkk}, the action (\ref{Mannact2}) can in fact be turned into a pure theory of gravity. We can accomplish this by redefining the metric according to $\tilde{g}_{\mu\nu}=e^{\phi/2}g_{\mu\nu}$, we may rewrite (\ref{Mannact2}) (dropping the constant and the matter contribution) as
\beq I_{\phi}[\tilde{g}]=\int d^{2}x\sqrt{-\tilde{g}}[\phi\tilde{R}-V(\phi)]\;,\quad V(\phi)=2\Lambda_{0}e^{-\phi/2}\;.\label{ManNaseinsteindila}\eeq
In this form we can solve the dilaton field equation and substitute the result back into the action, resulting in (a peculiar) pure theory of gravity with respect to the metric $\tilde{g}_{\mu\nu}$:
\beq I_{\phi=\phi_{0}}[\tilde{g}]=\int d^{2}x\sqrt{-g}f(\tilde{R})\;,\eeq
with $f(x)=2x\left[1-\log\left(\frac{-x}{\Lambda_{0}}\right)\right]$. For de Sitter or AdS solutions this is odd, as the resulting gravitational equations of motion give
\beq \tilde{R}f'(\tilde{R})-f(\tilde{R})=0\Rightarrow\tilde{R}=0\Rightarrow\tilde{g}_{\mu\nu}=0\;.\eeq
This means that the theory, with respect to $\tilde{g}$, does not admit (A)dS solutions. Of course, if we don't work in the $\tilde{g}$ frame, then one can study (A)dS solutions, as done in, \emph{e.g.,} \cite{Frassino:2015oca}.


\section{$D\to4$ Limit of Einstein-Gauss-Bonnet Gravity} \label{sec:D4limEGB}
\indent

Following the spirit of \cite{Mann:1992ar}, let's now try a similar trick with Einstein-Gauss-Bonnet gravity. This was recently accomplished in \cite{Fernandes:2020nbq,Hennigar:2020lsl}. Start with Einstein-Gauss-Bonnet gravity in $D$- dimensions, dropping any matter action,
\beq I_{EGB}=\int d^{D}x\sqrt{-g}\left[\frac{R}{16\pi G_{D}}-2\Lambda_{0}+\alpha_{GB}\mathcal{L}_{GB}\right]\;.\eeq
Now we subtract a term $\int d^{D}x\sqrt{-\tilde{g}}\alpha_{GB}\tilde{\mathcal{L}}_{GB}$, such that this term becomes a total derivative in $D=4$ dimensions, \emph{i.e., } $\lim_{D\to4}\alpha_{GB}\tilde{\mathcal{H}}_{\mu\nu}=0$. 

\begin{widetext}
Moreover, the metric $\tilde{g}_{\mu\nu}$ is conformally related to $g_{\mu\nu}$ via $\tilde{g}_{\mu\nu}=e^{\Phi}g_{\mu\nu}$, with $\sqrt{-\tilde{g}}=e^{D\Phi/2}\sqrt{-g}$, and the Gauss-Bonnet Lagrangian transforms\footnote{Reference \cite{Dabrowski:2008kx} writes their transformations a bit differently. For us, recall that $\Box e^{\Phi/2}=\frac{1}{2}e^{\Phi/2}\left(\Box\Phi+\frac{1}{2}(\nabla\Phi)^{2}\right)$ and $\nabla_{\mu}\nabla_{\nu}e^{\Phi/2}=\frac{1}{2}e^{\Phi/2}\left(\nabla_{\mu}\nabla_{\nu}\Phi+\frac{1}{2}(\nabla_{\mu}\Phi)(\nabla_{\nu}\Phi)\right)\;.$
Our expression for $\tilde{\mathcal{L}}_{GB}$ can be shown to be equivalent to Eq. (A5) in \cite{Hennigar:2020fkv}.} 
as \cite{Dabrowski:2008kx}:
\beq
\begin{split}
&\tilde{\mathcal{L}}_{GB}=e^{-4\Phi/2}\biggr\{\mathcal{L}_{GB}+4(D-3)G_{\mu\nu}\left(\nabla^{\mu}\nabla^{\nu}\Phi\right)+(D-3)(D-4)G_{\mu\nu}(\nabla^{\mu}\Phi)(\nabla^{\nu}\Phi)\\
&+(D-3)(D-2)\nabla_{\mu}[(\nabla^{\mu}\Phi)\Box\Phi-(\nabla_{\nu}\Phi)(\nabla^{\mu}\nabla^{\nu}\Phi)]\\
&+(D-2)(D-3)\biggr[\frac{(D-4)}{2}\Box\Phi(\nabla\Phi)^{2}+\frac{1}{2}\nabla_{\mu}[(\nabla\Phi)^{2}\nabla^{\mu}\Phi]\biggr]\\
&+\frac{1}{16}(D-1)(D-2)(D-3)(D-4)(\nabla\Phi)^{4}\biggr\}\;,
\end{split}
\label{conftLGB}
\eeq
where we used $\nabla_{\mu}\nabla^{\nu}(\nabla^{\mu}\Phi)=R^{\nu}_{\;\mu}\nabla^{\mu}\Phi+\nabla^{\nu}(\Box\Phi)$ and
\beq
\begin{split}
 (\Box\Phi)^{2}-(\nabla_{\mu}\nabla_{\nu}\Phi)^{2}&=(\nabla_{\nu}\Phi)(R^{\nu}_{\;\mu}\nabla^{\mu}\Phi)+\nabla^{\nu}[(\nabla_{\nu}\Phi)\Box\Phi]-\nabla_{\mu}[(\nabla_{\nu}\Phi)(\nabla^{\mu}\nabla^{\nu}\Phi)]\;.
\end{split}\eeq

Next, consider the action
\beq
\begin{split}
 I&=\int d^{D}x\sqrt{-g}\biggr(\frac{1}{16\pi G_{D}}(R-2\Lambda_{0})+\alpha_{GB}\left[\mathcal{L}_{GB}-\frac{\sqrt{-\tilde{g}}}{\sqrt{-g}}\tilde{\mathcal{L}}_{GB}\right]\biggr)\;.
\end{split}
\eeq
We now substitute in our conformally transformed Gauss-Bonnet Lagrangian density (\ref{conftLGB}), expanding our exponential factor to linear order in $O(D-4)$, 
\beq 
\begin{split}
I&=\int d^{D}x\sqrt{-g}\biggr[\frac{1}{16\pi G_{D}}(R-2\Lambda_{0})-\alpha_{GB}\biggr\{4(D-3)G_{\mu\nu}\left(\nabla^{\mu}\nabla^{\nu}\Phi\right)+(D-3)(D-4)G_{\mu\nu}(\nabla^{\mu}\Phi)(\nabla^{\nu}\Phi)\\
&+(D-3)(D-2)\nabla_{\mu}[(\nabla^{\mu}\Phi)\Box\Phi-(\nabla_{\nu}\Phi)(\nabla^{\mu}\nabla^{\nu}\Phi)]+(D-2)(D-3)\left[\frac{(D-4)}{2}\Box\Phi(\nabla\Phi)^{2}+\frac{1}{2}\nabla_{\mu}[(\nabla\Phi)^{2}\nabla^{\mu}\Phi]\right]\\
&+\frac{1}{16}(D-1)(D-2)(D-3)(D-4)(\nabla\Phi)^{4}\biggr\}-\alpha_{GB}\frac{(D-4)\Phi}{2}\biggr\{\mathcal{L}_{GB}+4(D-3)G_{\mu\nu}\left(\nabla^{\mu}\nabla^{\nu}\Phi\right)\\
&+(D-3)(D-2)\left(\nabla_{\mu}[(\nabla^{\mu}\Phi)\Box\Phi-(\nabla_{\nu}\Phi)(\nabla^{\mu}\nabla^{\nu}\Phi)]\right)+\frac{1}{2}(D-2)(D-3)\nabla_{\mu}[(\nabla\Phi)^{2}\nabla^{\mu}\Phi]\biggr\}\biggr]\;.
\end{split}
\label{startingeqappA}\eeq
Dropping all total derivatives,
\beq 
\begin{split}
I&=\int d^{D}x\sqrt{-g}\biggr[\frac{1}{16\pi G_{D}}(R-2\Lambda_{0})-\frac{\alpha_{GB}}{2}(D-4)\biggr\{\Phi\mathcal{L}_{GB}-2(D-3)G_{\mu\nu}(\nabla^{\mu}\Phi)(\nabla^{\nu}\Phi)\\
&+\frac{1}{2}(D-2)(D-3)(D-5)(\nabla\Phi)^{2}\Box\Phi+\frac{1}{8}(D-2)(D-3)(D-5)(\nabla\Phi)^{4}\biggr\}\biggr]\;.
\end{split}
\label{actionan3}\eeq
where we used
\beq \Phi G_{\mu\nu}\nabla^{\mu}\nabla^{\nu}\Phi=\nabla^{\mu}[\Phi G_{\mu\nu}\nabla^{\nu}\Phi]-G_{\mu\nu}(\nabla^{\mu}\Phi)(\nabla^{\nu}\Phi)\;,\eeq
\beq \Phi\nabla_{\mu}\left[(\nabla^{\mu}\Phi)\Box\Phi-(\nabla_{\nu}\Phi)(\nabla^{\mu}\nabla^{\nu}\Phi)\right]=\text{Tot. Der.}+(\nabla_{\mu}\Phi)(\nabla_{\nu}\Phi)(\nabla^{\mu}\nabla^{\nu}\Phi)-(\nabla\Phi)^{2}\Box\Phi\;,\eeq
and
\beq \Phi\nabla_{\mu}[(\nabla\Phi)^{2}(\nabla^{\mu}\Phi)]=\nabla_{\mu}[\Phi(\nabla\Phi)^{2}(\nabla^{\mu}\Phi)]-(\nabla\Phi)^{4}\;.\eeq
\end{widetext}

Now we invoke $\alpha_{GB}\to\alpha_{GB}/(D-4)$. In so doing, we see that we have eliminated all terms divergent in $1/(D-4)$, and remaining action is well defined in the $D=4$ limit,
\beq 
\begin{split}
I&=\int d^{4}x\sqrt{-g}\biggr[\frac{1}{16\pi G_{4}}(R-2\Lambda_{0})-\frac{\alpha_{GB}}{2}\biggr\{\Phi\mathcal{L}_{GB}\\
&-2G_{\mu\nu}(\nabla^{\mu}\Phi)(\nabla^{\nu}\Phi)-(\nabla\Phi)^{2}\Box\Phi-\frac{1}{4}(\nabla\Phi)^{4}\biggr\}\biggr]\;,
\end{split}
\label{actionan4}\eeq
matching the action given in \cite{Hennigar:2020lsl}. Since the Gauss-Bonnet Lagrangian by itself  remains purely topological in $D=4$, we see that \emph{locally} our theory has a constant shift symmetry in $\Phi$.

As pointed out in \cite{Hennigar:2020lsl}, when we make the following rescalings, 
\beq \Phi\to-2\Phi\;\quad g_{\mu\nu}\to-\frac{1}{2}g_{\mu\nu}\;,\quad \alpha_{GB}\to-\frac{\alpha_{GB}}{2}\;,\eeq
to our action (\ref{actionan4}) , we exactly reproduce the $D\to4$ limit of Einstein gravity plus the Kaluza-Klein reduced  Gauss-Bonnet Lagrangian over a flat internal space\footnote{The Kaluza-Klein metric ansatz is $ds^{2}_{D}=ds_{d}^{2}+e^{2\Phi}d\Sigma_{D-d}^{2}$, with $d\Sigma_{D-d}^{2}$ as the line element of the internal maximally symmetric space, while $\Phi$ is the metric function which depends on the external $d$-dimensional coordinates. Our action (\ref{KKredact}) is the limit where the internal space is flat. More generally, (\ref{KKredact}) is supplemented with a contribution which is non-vanishing for curved maximally symmetric internal space,
$$I_{\lambda}=\int d^{4}x\sqrt{-g}\left[-2\lambda Re^{-2\Phi}-12\lambda(\nabla\Phi)^{2}e^{-2\Phi}-6\lambda^{2}e^{-4\Phi}\right],$$
where $\lambda$ denotes the curvature of the internal space, \emph{e.g.,} $\lambda=0$ for vanishing curvature.}, written out explicitly in \cite{Lu:2020iav}:
\beq
\begin{split}
 I&=\int d^{4}x\sqrt{-g}\biggr[\frac{1}{16\pi G_{4}}(R-2\Lambda_{0})+\alpha_{GB}(\Phi\mathcal{L}_{GB}\\
&+4G^{\mu\nu}(\nabla_{\mu}\Phi)(\nabla_{\nu}\Phi)-4(\nabla\Phi)^{2}\Box\Phi+2(\nabla\Phi)^{4})\biggr]\;.
\end{split}
\label{KKredact}
\eeq

The $D\to4$ EGB model (\ref{actionan4}) or (\ref{KKredact}) describes an Einstein-Gauss-Bonnet theory of gravity coupled to a scalar field, analogous to the $D\to2$ limit of general relativity (\ref{Mannact2}). Unlike the $D\to2$ model of GR, we cannot perform a conformal rescaling on $g_{\mu\nu}$ and rewrite our $D\to4$ EGB theory as a pure theory of gravity. Moreover, we emphasize that our scalar-tensor theory is a special case of a Horndeski theory of gravity \cite{Horndeski:1974wa,Deffayet:2009mn,Charmousis12-1}. This is not surprising, as it is by now well known that dimensionally reducing Lovelock actions give the covariant, scalar-tensor theories of gravity with second order of equations motion \cite{VanAcoleyen11-1}. Indeed, the equations of motion of (\ref{KKredact}) are explicitly second order: \cite{Hennigar:2020lsl}
\beq
\begin{split}
0&=\frac{\alpha_{GB}}{2}\biggr\{-\mathcal{L}_{GB}+8G^{\mu\nu}\nabla_{\mu}\nabla_{\nu}\Phi+8R^{\mu\nu}(\nabla_{\mu}\Phi)(\nabla_{\nu}\Phi)\\
&-8(\Box\Phi)^{2}+8(\nabla\Phi)^{2}\Box\Phi+8(\nabla_{\mu}\nabla_{\nu}\Phi)^{2}\\
&+16(\nabla_{\mu}\Phi)(\nabla_{\nu}\Phi)(\nabla^{\mu}\nabla^{\nu}\Phi)\biggr\}
\end{split}
\label{phieom}\eeq
for $\Phi$, while for $g_{\mu\nu}$,
\beq
\begin{split}
0&=\frac{1}{8\pi G_{4}}(G_{\mu\nu}+\Lambda_{0}g_{\mu\nu})+\alpha_{GB}\biggr\{\mathcal{H}_{\mu\nu}-2R[(\nabla_{\mu}\Phi)(\nabla_{\nu}\Phi)\\
&+\nabla_{\mu}\nabla_{\nu}\Phi]+8R^{\delta}_{(\mu}\nabla_{\nu)}\nabla_{\delta}\Phi+8R^{\delta}_{(\mu}(\nabla_{\nu)}\Phi)(\nabla_{\delta}\Phi)\\
&-2G_{\mu\nu}[(\nabla\Phi)^{2}+2\Box\Phi]-4[(\nabla_{\mu}\Phi)(\nabla_{\nu}\Phi)+\nabla_{\mu}\nabla_{\nu}\Phi]\Box\Phi\\
&-[g_{\mu\nu}(\nabla\Phi)^{2}-4(\nabla_{\mu}\Phi)(\nabla_{\nu}\Phi)](\nabla\Phi)^{2}\\
&+8(\nabla_{(\mu}\Phi)(\nabla_{\nu)}\nabla_{\delta}\Phi)\nabla^{\delta}\Phi\\
&-4g_{\mu\nu}R^{\delta\rho}[\nabla_{\delta}\nabla_{\rho}\Phi+(\nabla_{\delta}\Phi)(\nabla_{\rho}\Phi)]+2g_{\mu\nu}(\Box\Phi)^{2}\\
&-2g_{\mu\nu}(\nabla_{\delta}\nabla_{\rho}\Phi)(\nabla^{\delta}\nabla^{\rho}\Phi)\\
&-4g_{\mu\nu}(\nabla^{\delta}\Phi)(\nabla^{\rho}\Phi)(\nabla_{\delta}\nabla_{\rho}\Phi)+4(\nabla_{\delta}\nabla_{\nu}\Phi)(\nabla^{\delta}\nabla_{\mu}\Phi)\\
&+4R_{\mu\delta\nu\rho}[(\nabla^{\delta}\Phi)(\nabla^{\rho}\Phi)+\nabla^{\rho}\nabla^{\delta}\Phi]\biggr\}\;,
\end{split}
\label{meteom}\eeq
 where $\mathcal{H}_{\mu\nu}$ is given in (\ref{partofEGBeom}). 

From the Kaluza-Klein reduction perspective, the equations of motion of (\ref{KKredact})   had to be second order since the scalar field $\Phi$ is simply a component of the full $D$-dimensional metric, and the starting $D$-dimensional action is EGB, which already has second order equations of motion. Similarly, from the Mann-Ross dimensionally regularized viewpoint, the equations of motion of (\ref{actionan4}) are second order since we are simply performing a conformal rescaling of EGB. 

Taking the trace of (\ref{phieom}) and (\ref{meteom}) and adding them together, we find
\beq 0=\frac{1}{8\pi G_{4}}(4\Lambda-R)-\frac{\alpha_{GB}}{2}\mathcal{L}_{GB}\;.\eeq
This should be compared to (\ref{traceEGB1}), for which we see that now the Gauss-Bonnet term will alter the local dynamics of spacetime in four-dimensions. As such, the theory is expected to have some non-trivial and interesting solutions. Black hole solutions to this model were initially considered in \cite{Lu:2020iav}, and further explored in \cite{Hennigar:2020lsl}, where four-dimensional Gauss-Bonnet Taub-NUTs were also studied. A complete survey of black hole solutions has not yet been accomplished for this model. Moreover, as a Horndeski theory of gravity, this model is expected to have potentially rich cosmological behavior. We will return to this momentarily. 

Lastly, we should emphasize that, unlike the original `novel' $D=4$ EGB theory \cite{Glavan:2019inb}, this $D\to4$ limit of EGB does not suffer from having a singular action in the $D\to4$ limit, and avoids the `index problem'. Since it is a Horndeski theory of gravity, it is also expected to have a well-posed Hamiltonian formulation.


\section{`Novel' Einstein-Lovelock Gravity}\label{sec:novelLL}

As eluded to in the introduction, general relativity and Gauss-Bonnet gravity are part of a wider class of theories known as Lovelock gravity \cite{Lovelock:1971yv,Lovelock:1972vz}. Therefore, our dimensional regularization scheme, essentially the technique introduced by Mann and Ross \cite{Mann:1992ar}, can be applied equally to more general Einstein-Lovelock actions. This was noted previously in \cite{Casalino:2020kbt}, however was accomplished only using the rescaling of the Lovelock couplings analogous to \cite{Glavan:2019inb}, and is thus expected to suffer from the issues mentioned earlier. 

 Let us briefly describe how the regularization scheme would work. Consider any Einstein-Lovelock theory of gravity, with action
\beq I_{ELL}=\int d^{D}x\sqrt{-g}\left[\frac{1}{16\pi G_{D}}(R-2\Lambda_{0})+\mathcal{L}_{LL}+\mathcal{L}_{\text{mat}}\right]\;,\label{ELLactionD}\eeq
where the Lovelock Lagrangian is 
\beq \mathcal{L}_{LL}=\sum_{n=0}^{t}\mathcal{L}_{(n)}=\sqrt{-g}\sum_{n=0}^{t}\alpha_{n}\mathcal{R}_{(n)}\eeq
with
\beq \mathcal{R}_{(n)}=\frac{1}{2^{n}}\delta^{\mu_{1}\nu_{1}...\mu_{n}\nu_{n}}_{\alpha_{1}\beta_{1}...\alpha_{n}\beta_{n}}\prod_{r=1}^{n}R^{\alpha_{r}\beta_{r}}_{\;\;\;\;\;\;\;\mu_{r}\nu_{r}}\;,\eeq
where $\alpha_{n}$ as the Lovelock coupling constants. Here we have used the generalized Kronecker delta symbol
\beq \delta^{\mu_{1}\nu_{1}...\mu_{n}\nu_{n}}_{\alpha_{1}\beta_{1}...\alpha_{n}\beta_{n}}=n!\delta^{\mu_{1}}_{[\alpha_{1}}\delta^{\nu_{1}}_{\beta_{1}}...\delta^{\mu_{n}}_{\alpha_{n}}\delta^{\nu_{n}}_{\beta_{n}]}\;.\eeq
The first few terms are explicitly,
\beq 
\mathcal{L}_{(0)}=\sqrt{-g}\alpha_{0}\;,\quad \mathcal{L}_{(1)}=\sqrt{-g}\alpha_{1}R\;,\quad \mathcal{L}_{(2)}=\mathcal{L}_{GB}\;,\eeq
\beq
\begin{split}
\mathcal{L}_{(3)}&=\sqrt{-g}\alpha_{3}\biggr\{R^{3}-12 RR_{\mu\nu}^{2}+16R_{\mu\nu}R^{\mu}_{\;\rho}R^{\nu\rho}\\
&+24R_{\mu\nu}R_{\rho\sigma}R^{\mu\rho\nu\sigma}+3RR_{\mu\nu\rho\sigma}^{2}\\
&-24R_{\mu\nu}R^{\mu}_{\;\rho\sigma\kappa}R^{\nu\rho\sigma\kappa}+4R_{\mu\nu\rho\sigma}R^{\mu\nu\eta\zeta}R^{\rho\sigma}_{\;\;\;\eta\zeta}\\
&-8R_{\mu\rho\nu\sigma}R^{\mu\;\;\;\nu}_{\;\eta\;\;\;\;\zeta}R^{\rho\eta\sigma\zeta}\biggr\}\;.
\end{split}
\eeq
By Einstein-Lovelock theories of gravity, we really mean to start our Lovelock Lagrangian at $n=2$. For even $D$, $t=D/2$, while for odd $D$, $t=(D-1)/2$. When $2n>D$, the quantity $\sqrt{-g}\mathcal{R}_{(n)}$ is the generalized Euler-density in $2n$-dimensions, 
\beq \chi(\mathcal{M}_{2n})=\frac{1}{(4\pi)^{n}n!}\int_{\mathcal{M}_{2n}}d^{2n}x\sqrt{-g}\mathcal{R}_{(n)}\;.\eeq
This tells us that $\mathcal{L}_{(n)}$ is topological in $2n$-dimensional spacetimes, while for $D>2n$, $\mathcal{L}_{(n)}$ contributes to local dynamics. 

Indeed, the field equations for pure Lovelock are 
\beq \mathcal{G}_{\mu\nu}=\sum_{n=0}^{t}\alpha_{n}\mathcal{G}^{(n)}_{\mu\nu}=\frac{1}{2}T_{\mu\nu}\;,\eeq
with
\beq \mathcal{G}^{(n)\alpha}_{\;\beta}=-\frac{1}{2^{n+1}}\delta^{\alpha\mu_{1}\nu_{1}...\mu_{n}\nu_{n}}_{\beta\sigma_{1}\rho_{1}...\sigma_{n}\rho_{n}}\prod_{p=1}^{n}R^{\mu_{p}\nu_{p}}_{\;\;\;\;\;\;\;\sigma_{p}\rho_{p}}\;,\eeq
which will vanish identically for $D\leq 2n$, due to the totally antisymmetric Kronecker delta symbol.

Following the insight of \cite{Glavan:2019inb}, it is expected that the Lovelock Lagrangian densities can contribute to local dynamics in $D=2n$ if we shift the Lovelock couplings via
\beq \alpha_{n}\to(D-2n)\alpha_{n}\;.\eeq

Our `novel' Einstein-Lovelock theory is then constructed in an analogous manner to our Einstein-Gauss-Bonnet theory of gravity. We start with the Einstein-Lovelock action (\ref{ELLactionD}), and subtract from it the highest dimensional term in $\tilde{\mathcal{L}}_{LL}$ with $\sqrt{\tilde{g}}$, where as before $\tilde{g}_{\mu\nu}=e^{\Phi}g_{\mu\nu}$, such that 
\beq\tilde{\mathcal{R}}_{(n)}=\frac{1}{2^{n}}\delta^{\mu_{1}\nu_{1}...\mu_{n}\nu_{n}}_{\alpha_{1}\beta_{1}...\alpha_{n}\beta_{n}}\prod_{r=1}^{n}\tilde{R}^{\alpha_{r}\beta_{r}}_{\;\;\;\;\;\;\;\mu_{r}\nu_{r}}\; ,\eeq
 with
\begin{widetext}
\beq
\begin{split}
&\tilde{R}_{\mu\nu\delta\rho}=e^{\Phi}\biggr[R_{\mu\nu\delta\rho}+\frac{1}{4}g_{\nu\rho}(\nabla_{\mu}\Phi)(\nabla_{\delta}\Phi)-\frac{1}{4}g_{\mu\rho}(\nabla_{\nu}\Phi)(\nabla_{\delta}\Phi)-\frac{1}{2}g_{\nu\rho}\nabla_{\mu}\nabla_{\delta}\Phi
+\frac{1}{2}g_{\mu\rho}\nabla_{\nu}\nabla_{\delta}\Phi\\
&-\frac{1}{4}g_{\nu\delta}(\nabla_{\mu}\Phi)(\nabla_{\rho}\Phi)+\frac{1}{4}g_{\mu\rho}(\nabla_{\nu}\Phi)(\nabla_{\rho}\Phi)+\frac{1}{2}g_{\nu\rho}\nabla_{\mu}\nabla_{\rho}\Phi-\frac{1}{2}g_{\mu\delta}\nabla_{\nu}\nabla_{\rho}\Phi+\frac{1}{4}(g_{\mu\rho}g_{\nu\delta}-g_{\mu\delta}g_{\nu\rho})(\nabla\Phi)^{2}\biggr]\; .
\end{split}
\eeq
\end{widetext}

  We then Taylor expand the subtracted action to terms only linear in $(D-2n)$, and rescale $\alpha_{n}\to (D-2n)\alpha_{n}$. For example, for cubic Lovelock gravity ($n=3$) we would write
\beq I^{(reg)}_{ELL}=I_{ELL}-\int d^{D}x\sqrt{\tilde{g}}\tilde{\mathcal{L}}_{(3)}\;,\eeq
where $\tilde{g}_{\mu\nu}$ and $\tilde{\mathcal{L}}_{(3)}$ are the conformally transformed metric and Lovelock Lagrangian density, respectively. We then Taylor expand the subtracted action to terms only linear in $(D-6)$, and rescale $\alpha_{3}\to (D-6)\alpha_{3}$. We may then safely take the $D=6$ limit, and study a novel cubic Lovelock gravity in $D=6$. 

As our EGB theory resulted in Horndeski gravity upon our method of `dimensional regularization' it is expected that our novel Einstein-Lovelock theory will describe generalized Horndeski theories of gravity \cite{Deffayet:2009mn}, which to leading order arise from dimensional reduction of Lovelock theories of gravity \cite{VanAcoleyen11-1}.


\section{Dimensionally Regularized Effective String Action}\label{sec:effstring}

One of the fascinating features of string theory is that Einstein's equations emerge through the requirement of avoiding a conformal anomaly, \emph{ i.e.}, maintaining conformal invariance of a quantized string on a curved background. Just as in quantum field theory, one will attempt to impose a classical symmetry, \emph{e.g.}, conformal invariance, at various orders of the loop expansion in the action. In string theory there are two such perturbative expansions -- the $\alpha'$ expansion, one which is unique to string theory, and the genus expansion, the 2-D analog of performing a Feynman diagram analysis. It is the conformal invariance of the $\alpha'=\lambda^{2}_{s}/2\pi$ corrections which gives rise to Einstein's equations and its higher curvature corrections. In particular, at \emph{tree-level} ($\alpha'=0$), imposing conformal invariance requires $R_{\mu\nu}=0$ -- Einstein's vacuum equations. 

 At higher order $\alpha'$ corrections, demanding conformal symmetry at the quantum level modifies Einstein's equations, in which one is forced to include additional fields as well as higher curvature terms \cite{Zwiebach:1985uq,Metsaev87-1,Green12-1}. By demanding conformal invariance, the series of $\alpha'$ corrections is constrained due to the condition of conformal invariance as applied to scattering amplitudes in the string S-matrix. At each level of approximation, there is an intrinsic ambiguity due to field redefinitions which preserve general covariance and gauge invariance. These will give rise to a variety of different actions, each, however, having an equivalent S-matrix. A particular such action is the gravi-dilaton sector of the 10-dimensional heterotic superstring  \cite{Zwiebach:1985uq,Metsaev87-1}; or here in general dimension $D$ \cite{Gasperini07-1}
\beq 
\begin{split}
I&=-\frac{1}{2\lambda^{D-2}_{s}}\int d^{D}x\sqrt{-g}e^{-\phi}\biggr[R+(\nabla\phi)^{2}\\
&-\frac{\alpha'}{4}\mathcal{L}_{GB}+\frac{\alpha'}{4}(\nabla\phi)^{4}\biggr]\;,
\end{split}
\label{effstringact1}\eeq
where $\lambda_{s}$ is a constant with units of length, representing the characteristic string length in string theory. Note that in the limit $D\to4$ the Gauss-Bonnet piece contributes due to the presence of the dilaton factor $e^{-\phi}$.

Earlier we described a `novel' Einstein-Lovelock theory via a type of dimensional regularization. Pure theories of Lovelock gravity, however, are not known to exist in nature. Higher derivative Lovelock contributions, as seen above, naturally arise in string theory from $\alpha'$ perturbative corrections to the string. Thus it is natural to apply our dimensional regularization to a low energy effective string action and take the four dimensional limit as described above\footnote{Following \cite{VanAcoleyen11-1}, the model (\ref{effstringact1}) provides us with an action upon which a Kaluza-Klein dimensional reduction leads to a Horndeski theory of gravity, such that covariant galileons have a natural stringy origin. This is the subject of unpublished work \cite{TMA20}.
}.

In particular, we will consider the low-energy effective string action (\ref{effstringact1}), and subtract from it 
\beq \tilde{I}_{1}=-\frac{1}{2\lambda_{s}^{d-1}}\int d^{D}x\sqrt{-\tilde{g}}e^{-\phi}\left(-\frac{\alpha'}{4}\tilde{\mathcal{L}}_{GB}+\frac{\alpha'}{4}(\tilde{\nabla}\phi)^{4}\right)\;,\eeq
where quantities with a $\tilde{.}$ correspond to geometric objects in the conformally related background $\tilde{g}_{\mu\nu}=e^{\Phi}g_{\mu\nu}$, and here we are writing the dilaton as $\phi$ to avoid confusion with scalar field $\Phi$ relating the conformally transformed metric. 

We should emphasize that unlike the pure theories of gravity we have considered thus far, our `regularizing' action $\tilde{I}_{1}$ does \emph{not} become a total derivative in the limit $D\to4$. This is because, as in the effective string action (\ref{effstringact1}), the Gauss-Bonnet Lagrangian is multiplied by a dilaton factor $e^{-\phi}$. Consequently, the dimensionally regularized theory we are proposing (\ref{Iregst}) is different from the `novel' Lovelock theories of gravity discussed earlier in that $I_{reg}$ (\ref{Iregst}) is not the same theory as (\ref{effstringact1}) even in the $D=4$ limit\footnote{We would like to thank an anonymous referee for encouraging us to emphasize this relevant point.}. Nonetheless, we will find that our dimensionally regularized action leads to interesting new features not observed by the effective action (\ref{effstringact1}) in the $D\to4$ limit, as discussed below. 

Then, 
\begin{widetext}
\beq
\begin{split}
I_{reg}&=I-\tilde{I}_{1}=-\frac{1}{2\lambda^{D-2}_{s}}\int d^{D}x\sqrt{-g}e^{-\phi}\biggr[R-2\Lambda_{0}+(\nabla\phi)^{2}-\frac{\alpha'}{4}\left(\mathcal{L}_{GB}-\frac{\sqrt{-\tilde{g}}}{\sqrt{-g}}\tilde{\mathcal{L}}_{GB}\right)\\
&+\frac{\alpha'}{4}\left((\nabla\phi)^{4}-\frac{\sqrt{-\tilde{g}}}{\sqrt{-g}}(\tilde{\nabla}\phi)^{4}\right)\biggr]\;,
\end{split}
\label{Iregst}\eeq
where we have explicitly included a cosmological constant term $\Lambda_{0}$. Using our conformally transformed Gauss-Bonnet Lagrangian density (\ref{conftLGB}), we have as before
\beq
\begin{split}
&\left(\mathcal{L}_{GB}-\frac{\sqrt{-\tilde{g}}}{\sqrt{-g}}\tilde{\mathcal{L}}_{GB}\right)=\mathcal{L}_{GB}-e^{(D-4)\Phi/2}\biggr\{\mathcal{L}_{GB}+4(D-3)G_{\mu\nu}\left(\nabla^{\mu}\nabla^{\nu}\Phi\right)\\
&+(D-3)(D-4)G_{\mu\nu}(\nabla^{\mu}\Phi)(\nabla^{\nu}\Phi)+(D-3)(D-2)\nabla_{\mu}[(\nabla^{\mu}\Phi)\Box\Phi-(\nabla_{\nu}\Phi)(\nabla^{\mu}\nabla^{\nu}\Phi)]\\
&+(D-2)(D-3)\left[\frac{(D-4)}{2}\Box\Phi(\nabla\Phi)^{2}+\frac{1}{2}\nabla_{\mu}[(\nabla\Phi)^{2}\nabla^{\mu}\Phi]\right]+\frac{1}{16}(D-1)(D-2)(D-3)(D-4)(\nabla\Phi)^{4}\biggr\}\;.
\end{split}
\eeq
\end{widetext}
Also, using $\tilde{\nabla}_{\mu}\phi=\nabla_{\mu}\phi$ so 
\beq \tilde{\nabla}^{\mu}\phi=e^{-\Phi}\nabla^{\mu}\phi\Rightarrow (\tilde{\nabla}\phi)^{4}\to e^{-4\Phi/2}(\nabla\phi)^{4}\;,\eeq
we have
\beq \left((\nabla\phi)^{4}-\frac{\sqrt{-\tilde{g}}}{\sqrt{-g}}(\tilde{\nabla}\phi)^{4}\right)=(\nabla\phi)^{4}\left(1-e^{(D-4)\frac{\Phi}{2}}\right)\;.\eeq

The next step is to expand the exponential factor $e^{(D-4)\frac{\Phi}{2}}\approx 1+(D-4)\frac{\Phi}{2}+O((D-4)^{2})$, and then rescale $\alpha'\to\alpha'/(D-4)$. We immediately see a problem, however. Unlike the pure EGB case, the overall dilaton factor $e^{-\phi}$ spoils our ability to drop total derivatives, such that our rescaling of $\alpha'$ and then the limit $D\to4$ leads to an action with a divergence. For example, upon expanding  $e^{(D-4)\frac{\Phi}{2}}$, we have at leading order a term proportional to
\beq 
\begin{split}
\alpha'e^{-\phi}e^{(D-4)\frac{\Phi}{2}}G_{\mu\nu}(\nabla^{\mu}\nabla^{\nu}\Phi)&\approx\alpha'e^{-\phi}G_{\mu\nu}(\nabla^{\mu}\nabla^{\nu}\Phi)\\
&+\alpha'\mathcal{O}(D-4)+...\;.
\end{split}
\label{exterm1}\eeq
In the pure EGB case we did not have a dilaton contribution and so this was a total derivative which could be dropped from the action; had we been unable to drop this term, $\alpha'\to\alpha'/(D-4)$ would lead to a divergence in the action upon taking $D\to4$. Thus, the presence of a \emph{dynamical} dilaton $\phi$ spoils our ability to take the $D\to4$ limit. One way to avoid this problem and keep $\phi$ dynamical is to also rescale the dilaton $\phi$,
\beq  \phi\to(D-4)\phi\;.\label{rescaledilaton}\eeq
Then we can perform a Taylor series expansion and keep only those terms linear in $(D-4)$, such that (\ref{exterm1}) becomes:
\beq
\begin{split}
&\alpha'e^{(D-4)[\frac{\Phi}{2}-\phi]}G_{\mu\nu}(\nabla^{\mu}\nabla^{\nu}\Phi)\approx\alpha'G_{\mu\nu}(\nabla^{\mu}\nabla^{\nu}\Phi)\\
&+\alpha'(D-4)\left(\frac{\Phi}{2}-\phi\right)G_{\mu\nu}(\nabla^{\mu}\nabla^{\nu}\Phi)+...
\end{split}
\eeq
The first of these is simply a total derivative and can be removed from the action while the second, since it is proportional to $(D-4)$, will not become singular in the limit $D\to4$, after rescaling $\alpha'\to\alpha'/(D-4)$.

Note, moreover, that upon (\ref{rescaledilaton}), the self-interactions of the dilaton $\phi$ will vanish from the action upon the $D\to4$ limit. This is because $(\nabla\phi)^{p}\to(D-4)^{p}(\nabla\phi)^{p}\to0$ for $p\geq1$. Altogether then, our dimensionally regularized low-energy effective string action to order $O(D-4)$ is a simple modification of (\ref{startingeqappA}),
where upon taking the $\alpha'\to\alpha'/(D-4)$, and subsequently $D\to4$, we have the dimensionally regularized (\ref{actionan4}) plus a dilaton $\phi$ correction:
\begin{widetext}
\beq 
\begin{split}
I_{reg}&=-\frac{1}{2\lambda_{s}^{2}}\int d^{4}x\sqrt{-g}\biggr[R-2\Lambda_{0}-\frac{\alpha'}{8}\biggr(\Phi\mathcal{L}_{GB}-2G_{\mu\nu}(\nabla^{\mu}\Phi)(\nabla^{\nu}\Phi)-(\nabla\Phi)^{2}\Box\Phi-\frac{1}{4}(\nabla\Phi)^{4}\biggr)\\
&+\frac{\alpha'}{4}\phi\biggr(\mathcal{L}_{GB}+4G_{\mu\nu}\left(\nabla^{\mu}\nabla^{\nu}\Phi\right)+2\left(\nabla_{\mu}[(\nabla^{\mu}\Phi)\Box\Phi-(\nabla_{\nu}\Phi)(\nabla^{\mu}\nabla^{\nu}\Phi)]\right)+\nabla_{\mu}[(\nabla\Phi)^{2}\nabla^{\mu}\Phi]\biggr)\biggr]\;.
\end{split}
\label{actionstringd4}\eeq
\end{widetext}
We observe that because we have rescaled the dilaton (\ref{rescaledilaton}), we do not have an overall factor of $e^{-\phi}$, as one would have in the usual string frame.  So our dimensionally regularized action appears as a type of $D\to4$ limit of the low-energy effective string action (\ref{effstringact1}) in Einstein frame. 

Let's comment on comparing to the Kaluza-Klein dimensional reduction of (\ref{effstringact1}). Following \cite{VanAcoleyen11-1}, it is tedious but straightforward to dimensionally reduce (\ref{effstringact1}), where we must also reduce the $D$-dimensional dilaton $\phi(x)$. This is accomplished noting that it is perfectly consistent to truncate higher modes in the Kaluza-Klein tower such that $\phi$ depends only on the coordinates of the lower dimensional external space, and not on the coordinates of the compactified internal space. Upon compactification, the resulting four-dimensional string action includes kinetic terms for the dilaton, as well as an overall $e^{-\phi}$. Since these features are missing from our $D\to4$ limit of (\ref{effstringact1}), it seems unlikely that we will match with the Kaluza-Klein reduction of the effective string action, except for potentially special redefintions of $\phi$ and $\Phi$. It could be worthwhile to pursue this further. 

Interestingly, notice that we now have terms where the dilaton $\phi$ `interacts' wth the conformal rescaling $\Phi$, when $\phi$ is not a simple scalar multiple of $\Phi$. Thus, we arrive at a particular sub-sector of multi-Galileon (or multi-field generalization of Horndeski) gravity \cite{Padilla:2012dx,Kobayashi:2013ina}, albeit in a rather strange form.  Moreover, there are some additional special cases for $\phi$. First, if we set $\phi=0$ at this stage, we simply recover (\ref{actionan4}), as expected. Moreover, if we set $\phi=\Phi/2$, we have special Horndeski theory of gravity, without a Gauss-Bonnet contribution, 
\beq 
\begin{split}
I&=-\frac{1}{2\lambda^{2}_{s}}\int d^{4}x\sqrt{-g}\biggr[R-2\Lambda_{0}\\
&-\frac{\alpha'}{4}\biggr(G_{\mu\nu}(\nabla^{\mu}\Phi)(\nabla^{\nu}\Phi)+\Box\Phi(\nabla\Phi)^{2}+\frac{3}{8}(\nabla\Phi)^{4}\biggr)\biggr]\;.
\end{split}
\label{dimregstringqactionnodil}\eeq
It would be interesting to further study the properties of the action (\ref{actionstringd4}), including its solutions, however we will leave this for future work and move on to study cosmological solutions of the $D\to4$ limit of Einstein-Gauss-Bonnet gravity (\ref{actionan4}).


\section{Cosmological Solutions of `Novel' Einstein-Gauss-Bonnet} \label{sec:cosmo}
\indent

Here we will consider some elementary cosmological solutions of (\ref{phieom}) and (\ref{meteom}), focusing on the pure theory (\ref{KKredact}) without the addition of external matter\footnote{This is in contrast with \cite{Glavan:2019inb}, where the authors consider cosmological solutions after adding a canonical scalar field to the action.}. We point out that the cosmological solutions to the  action we consider (\ref{KKredact}), is lacking a kinetic term for $\Phi$ at the same level as the Ricci scalar (we have kinetic terms at the level of the Gauss-Bonnet contribution). Note, in fact, we can acquire a kinetic term by dimensionally regularizing the full EGB action, in which our final action would include a conformally rescaled Einstein-Hilbert contribution. However, as noted in \cite{Hennigar:2020lsl}, a field redefinition transforms this alternative regularized action into (\ref{KKredact}).

Without a kinetic term for $\Phi$, some might consider the cosmological solutions we arrive at to not be entirely realistic. This is a valid critique, and we will indeed uncover some peculiar aspects of our solutions; for more realistic phenomenology we should seek to modify our action. Nonetheless, the model we study here  can be considered to be the theory closest to a $D=4$ Einstein-Gauss-Bonnet theory of gravity, and so it is worthwhile studying its own solutions without modification. 

Taking the 4-dimensional spatially flat FLRW metric ansatz for an isotropic universe, 
\beq \label{frw}
ds^2=-dt^2+a(t)^2\delta_{ij}dx^i dx^j, \ \ \ \ \ \ \Phi=\Phi(t),
\eeq
we obtain the equation of motion for the scalar field
\beq \label{phiFRWeom}
0=24\tilde{\alpha}_{GB} a^3\Big(H-\Phi'\Big)^2\Big(H^2+H'-H\Phi'-\Phi''\Big),
\eeq
where $H=a'/a$ is the usual Hubble parameter and $\tilde{\alpha}_{GB}=16\pi G_4\alpha_{GB}$. The Friedmann constraint and Einstein equation become
\beq \label{fried}
\begin{split} 
0&=-2a^3\Big[ \Lambda_0-3H^2 \\
&-\tilde{\alpha}_{GB}\Big(12H^3\Phi'-18(H\Phi')^2+12H(\Phi')^3-3(\Phi')^4\Big)\Big]\;,
\end{split}
\eeq

\beq\label{stein}
\begin{split} 
0&=6a^2\Big[ 3H^2+2H'-\Lambda_0  +\tilde{\alpha}_{GB}\Big(8 H^3 \Phi '+4 H^2 \Phi ''\\
&-6 H^2 \left(\Phi '\right)^2 +8 H \Phi ' \left(H'-\Phi ''\right)\\
&+\left(\Phi '\right)^2 \Big(-4 H'+4 \Phi ''+\left(\Phi '\right)^2\Big)\Big)\Big]\;.     \end{split}
  \eeq

  \subsection{Constant curvature}

Interestingly, the equation of motion for the scalar field, (\ref{phiFRWeom}), factorizes. We can immediately read off one class of solutions by setting the first factor to zero, so that
\beq
\Phi'=H.
\eeq
Inserting the above into the Friedmann constraint (\ref{fried}), we obtain a fourth-order polynomial in the Hubble parameter:
\begin{equation}
    0=-2a^3\Big(\Lambda_0-3(H^2 + \tilde{\alpha}_{GB} H^4)\Big).
\end{equation}
There are four solutions to the above equation for $H$. Assuming a positive value for the cosmological constant, we find two solutions are real when $\tilde{\alpha}_{GB}>0$, while the other two are real for $\tilde{\alpha}_{GB} <0:$
\begin{equation}\label{goodHub}
    H_{\pm}^{\tilde{\alpha}_{GB} >0}=\pm\frac{1}{\sqrt{6}}\Bigg[\frac{\sqrt{3}\sqrt{3+4\tilde{\alpha}_{GB}\Lambda_0}}{\tilde{\alpha}_{GB}}-\frac{3}{\tilde{\alpha}_{GB}}\Bigg]^{1/2}, 
\end{equation}
\begin{equation}
    H_{\pm}^{\tilde{\alpha}_{GB} <0}=\pm\frac{1}{\sqrt{6}}\Bigg[-\frac{\sqrt{3}\sqrt{3+4\tilde{\alpha}_{GB}\Lambda_0}}{\tilde{\alpha}_{GB}}-\frac{3}{\tilde{\alpha}_{GB}}\Bigg]^{1/2}.
\end{equation}
These solutions also satisfy (\ref{stein}). For each of the four roots, the scalar field and scale factor are given by
\begin{equation}\label{genPhianda}
    \Phi_{\pm}(t)=H_{\pm}t+\Phi_0, \ \ \ \ \ \ \ \ \ a_{\pm}(t)=e^{H_{\pm}t},
\end{equation}
where $\Phi_0$ is an integration constant. Note that the following limits hold:
\begin{equation}\label{noCC}
    \begin{split}
        \lim_{\Lambda_0\rightarrow 0}H^{\tilde{\alpha}_{GB} >0}_{\pm}=0,  \ \ \ \ \ \lim_{\Lambda_0\rightarrow 0}H^{\tilde{\alpha}_{GB} < 0}_{\pm} = \pm\frac{1}{\sqrt{-\tilde{\alpha}_{GB}}},
    \end{split}
\end{equation}
\begin{equation}\label{noalpha}
    \lim_{\tilde{\alpha}_{GB}\rightarrow 0}H^{\tilde{\alpha}_{GB} >0}_{\pm}=\pm\sqrt{\frac{\Lambda_0 }{3}},  \ \ \ \ \ \ \lim_{\tilde{\alpha}_{GB} \rightarrow 0}H_{\pm}^{\tilde{\alpha}_{GB} <0}=\pm i_{\infty},
\end{equation}
where $i_{\infty}$ is imaginary infinity. 

As was mentioned in section \ref{sec:effstring}, one expects the Gauss-Bonnet coupling $\alpha_{GB}$ to be proportional to the string theory expansion parameter $\alpha'=\lambda_s^2/2\pi$. Since $\alpha'>0,$ consistency with string theory implies that we take $\alpha_{GB}>0.$ The limits (\ref{noCC}) and (\ref{noalpha}) support that conclusion. Consequently, we find that the theory (\ref{KKredact}) with $\Lambda_0\rightarrow0$ only emits Minkowski space as a constant curvature solution, compatible with the first limit in (\ref{noCC}), while the $\alpha_{GB}\rightarrow 0$ limit recovers usual (A)dS space with $H^2=\Lambda_0/3$ and no additional content. With both parameters turned on, we have the solution class for $\Phi$ being linear in cosmological time as in (\ref{genPhianda}), with the Hubble parameter given in (\ref{goodHub}).


\subsection{Late time power law}

If we instead take the ansatz $a(t)=(t/t_0)^p\equiv\tt^p,$ then the field equation for $\Phi(t)$ becomes
\begin{equation}\label{powerPhiEOM}
0= \Big(\frac{p}{\tt}-\Phi'\Big)^2\Big(\frac{p(p-1-\tt\Phi')}{\tt^2}-\Phi''\Big).
\end{equation}
The second factor can be integrated, obtaining
\begin{equation}\label{powerPhi}
\Phi=p\log\tt+c_0\frac{\tt^{1-p}}{1-p}+\Phi_0,
\end{equation}
where $c_0$ and $\Phi_0$ are constants. Note that the $c_0=0$ solution satisfies vanishing of the first factor in (\ref{powerPhiEOM}). (\ref{powerPhi}) forces the Friedmann and Einstein equations to be
\begin{equation}
0=\frac{\Lambda_0}{3}-\frac{p^2}{\tt^2}+\frac{\tilde{\alpha}_{GB}c_0^4}{\tt^{4p}}-\frac{\tilde{\alpha}_{GB}p^4}{\tt^4},
\end{equation}
\begin{equation}
0=    \Lambda_0+\frac{p(2-3p)}{\tt^2}-\frac{\tilde{\alpha}_{GB} c_0^4}{\tt^{4p}}+\frac{\tilde{\alpha}_{GB} p^3(4-3p)}{\tt^4}.
\end{equation}
For late times, we have
\begin{equation}\label{LTfried}
0=\frac{\Lambda_0}{3}-\frac{1}{\tt^2}\Big(p^2-\frac{\tilde{\alpha}_{GB}c_0^4}{\tt^{4p-2}}\Big)+O(\tt^{-4}),
\end{equation}
\begin{equation}\label{LTstein}
0=\Lambda_0+\frac{1}{\tt^2}\Big(p(2-3p)-\frac{\tilde{\alpha}_{GB}c_0^4}{\tt^{4p-2}}\Big)+O(\tt^{-4}),
\end{equation}
which illustrates that for power law solutions consistent with both (\ref{LTfried}) and (\ref{LTstein}), we must have that $\Lambda_0\rightarrow 0$ and $p=1/2.$ The terms in the large parenthesis in both expressions vanish, and the integration constant $c_0$ is fixed to be 
\begin{equation}\label{c0}
c_0^4=\frac{1}{4\tilde{\alpha}_{GB}},
\end{equation}
while the scale factor is
\begin{equation}\label{radiationdom}
a(t)=\sqrt{t/t_0},
\end{equation}
corresponding to radiation-like dominated expansion. The solution for $\Phi$ is given by (\ref{powerPhi}) with $c_0$ as in (\ref{c0}).  

Stability of the solutions is left for future work and can be quite intricate in theories with non-standard kinetic Galileon terms, as the existence of ghost and gradient instabilities and superluminal propagating modes can depend on any external matter backgound~\cite{Easson:2013bda}.
In general, cosmology involving non-standard Lagrangians of the Galileon type exhibit extremely rich cosmological structure, including accelerated inflationary expansion~\cite{Creminelli:2010ba,Kobayashi:2010cm} and bouncing models~\cite{Qiu:2011cy,Easson:2011zy}.  Lagrangians of this type have the intriguing feature of being able to violate the Null Energy Condition (NEC) in a stable way~\cite{Deffayet:2010qz,Creminelli:2010ba,Kobayashi:2010cm,Elder:2013gya,Rubakov:2014jja}.
We leave a more complete discussion of cosmological solutions in this model to future work. 


\section{Discussion} \label{sec:disc}
\indent

Recently a pure theory of Einstein-Gauss-Bonnet gravity was written down in four-spacetime dimensions \cite{Glavan:2019inb}, seemingly eluding Lovelock's theorem. It is unclear how well-posed this `novel' theory of gravity is. Here we applied a dimensional regularization technique, first used by Mann and Ross \cite{Mann:1992ar}, to the case of pure Einstein-Gauss-Bonnet gravity, resulting in a particular Horndeski theory of gravity matching the result found via a Kaluza-Klein reduction over a flat internal space \cite{Lu:2020iav}. This result was also recently reported by \cite{Fernandes:2020nbq,Hennigar:2020lsl} . We then commented on how to adapt the technique to higher curvature Lovelock theories of gravity, as well as  a low-energy effective string action with an $\alpha'$ correction. With respect to the $D\to4$ limit of the $\alpha'$-corrected string action, we found we must also rescale the dilaton $\phi$ to have a non-singular action in $D=4$, resulting in an action that does not have a self-kinetic term for the dilaton, unlike what is expected for a Kaluza-Klein reduction of the string action. Interestingly, when the conformal rescaling $\Phi$ is interpreted as another dilaton, the resulting action appears to be a special case of a covariant multi-Galileon theory of gravity. Lastly, we studied some elementary cosmological solutions to the Horndeski theory gravity emerging from the $D\to4$ limit of Einstein-Gauss-Bonnet, finding a class of constant curvature solutions for $\Phi$ being linear in cosmological time, and a late time radiation-like dominated expansion with zero cosmological constant and a somewhat nontrivial $\Phi.$

There are a number of new directions to take our work. First, it would be worthwhile to study a specific higher curvature Lovelock theory of gravity, \emph{e.g.}, one with cubic interactions, and see whether the Mann-Ross regularization technique would match a Kaluza-Klein reduction of the same model. This may lead to a new of understanding Kaluza-Klein reduction, or a better understanding of the dimensional regularization method employed here. Indeed, it seems this regularization technique would \emph{not} match the the Kaluza-Klein reduction for the effective string action. 

Studying the $D\to4$ limit of the effective string action would be interesting as it may lead to particular models of multi-Galileons with potentially interesting physics. Moreover, we found that particular choices of the dilaton $\phi$ lead to simpler scalar-tensor theories of gravity, which may have potentially interesting solutions. 

The $D\to4$ limit of Einstein-Gauss-Bonnet gravity reported here has already been analyzed in some detail, \emph{e.g.} Taub-NUT solutions in \cite{Hennigar:2020lsl} and AdS black holes in the $D\to3$ limit \cite{Hennigar:2020fkv}, but requires further study. Indeed, it would be interesting to consider other types of black hole solutions to the theory, as well as star-like objects. Moreover, we only scratched the surface of cosmological solutions to this model and these deserve further analysis, including studying the stability of these solutions, with and without external matter (along the lines of \cite{Easson:2013bda}), and also considering matter density perturbations.

 In fact, as a Horndeski theory of gravity, the $D\to4$ limit of EGB opens up a wide range of potentially interesting avenues. This includes studying more complicated models of inflation, like $G$-inflation \cite{Kobayashi:2010cm}, and traversable wormholes, as our specific Horndeski theory may evade the initial assumptions of the no-go theorem presented in \cite{Rubakov:2015gza,Rubakov:2016zah}. 

\vspace{-1mm}

Finally, the Vainshtein mechanism of mass screening, leading to a `fifth force' has been studied for a particular four dimensional model of Einstein-Gauss-Bonnet gravity, defined via a $D$-dimensional Kaluza-Klein reduction, \emph{e.g.}, \cite{Gannouji:2011qz}. Due to the similarities between the dimensionally regularized action and the model found using dimensional reduction, it would be interesting to examine the Vainshtein mechanism in the context of the `novel' EGB theory discussed here. We leave each of these research avenues for future work. 
 
\vspace{1mm}

\section*{ACKNOWLEDGMENTS}

\vspace{-1mm}

The work of DAE is supported in part by a grant from the Foundational Questions Institute (FQXi). We would like to thank an anonymous referee who provided insightful feedback which led to the improvement of the original version of this manuscript. Again for good measure, while this work was in preparation we became aware of the work \cite{Fernandes:2020nbq,Hennigar:2020lsl}, which has significant overlap with Section \ref{sec:D4limEGB}.


\bibliography{D4EGBrefs}

\begin{thebibliography}{66}
\expandafter\ifx\csname natexlab\endcsname\relax\def\natexlab#1{#1}\fi
\expandafter\ifx\csname bibnamefont\endcsname\relax
  \def\bibnamefont#1{#1}\fi
\expandafter\ifx\csname bibfnamefont\endcsname\relax
  \def\bibfnamefont#1{#1}\fi
\expandafter\ifx\csname citenamefont\endcsname\relax
  \def\citenamefont#1{#1}\fi
\expandafter\ifx\csname url\endcsname\relax
  \def\url#1{\texttt{#1}}\fi
\expandafter\ifx\csname urlprefix\endcsname\relax\def\urlprefix{URL }\fi
\providecommand{\bibinfo}[2]{#2}
\providecommand{\eprint}[2][]{\url{#2}}

\bibitem[{\citenamefont{Lovelock}(1971)}]{Lovelock:1971yv}
\bibinfo{author}{\bibfnamefont{D.}~\bibnamefont{Lovelock}},
  \bibinfo{journal}{J. Math. Phys.} \textbf{\bibinfo{volume}{12}},
  \bibinfo{pages}{498} (\bibinfo{year}{1971}).

\bibitem[{\citenamefont{Lovelock}(1972)}]{Lovelock:1972vz}
\bibinfo{author}{\bibfnamefont{D.}~\bibnamefont{Lovelock}},
  \bibinfo{journal}{J. Math. Phys.} \textbf{\bibinfo{volume}{13}},
  \bibinfo{pages}{874} (\bibinfo{year}{1972}).

\bibitem[{\citenamefont{Chatterjee and Parikh}(2014)}]{Chatterjee:2013daa}
\bibinfo{author}{\bibfnamefont{S.}~\bibnamefont{Chatterjee}} \bibnamefont{and}
  \bibinfo{author}{\bibfnamefont{M.}~\bibnamefont{Parikh}},
  \bibinfo{journal}{Class. Quant. Grav.} \textbf{\bibinfo{volume}{31}},
  \bibinfo{pages}{155007} (\bibinfo{year}{2014}), \eprint{1312.1323}.

\bibitem[{\citenamefont{Glavan and Lin}(2020)}]{Glavan:2019inb}
\bibinfo{author}{\bibfnamefont{D.~z.} \bibnamefont{Glavan}} \bibnamefont{and}
  \bibinfo{author}{\bibfnamefont{C.}~\bibnamefont{Lin}},
  \bibinfo{journal}{Phys.\ Rev.\ Lett.} \textbf{\bibinfo{volume}{124}},
  \bibinfo{pages}{081301} (\bibinfo{year}{2020}), \eprint{1905.03601}.

\bibitem[{\citenamefont{Casalino et~al.}(2020)\citenamefont{Casalino, Colleaux,
  Rinaldi, and Vicentini}}]{Casalino:2020kbt}
\bibinfo{author}{\bibfnamefont{A.}~\bibnamefont{Casalino}},
  \bibinfo{author}{\bibfnamefont{A.}~\bibnamefont{Colleaux}},
  \bibinfo{author}{\bibfnamefont{M.}~\bibnamefont{Rinaldi}}, \bibnamefont{and}
  \bibinfo{author}{\bibfnamefont{S.}~\bibnamefont{Vicentini}}
  (\bibinfo{year}{2020}), \eprint{2003.07068}.

\bibitem[{\citenamefont{Fernandes}(2020)}]{Fernandes:2020rpa}
\bibinfo{author}{\bibfnamefont{P.~G.} \bibnamefont{Fernandes}}
  (\bibinfo{year}{2020}), \eprint{2003.05491}.

\bibitem[{\citenamefont{Konoplya and Zhidenko}(2020)}]{Konoplya:2020qqh}
\bibinfo{author}{\bibfnamefont{R.}~\bibnamefont{Konoplya}} \bibnamefont{and}
  \bibinfo{author}{\bibfnamefont{A.}~\bibnamefont{Zhidenko}},
  \bibinfo{journal}{Phys. Rev. D} \textbf{\bibinfo{volume}{101}},
  \bibinfo{pages}{084038} (\bibinfo{year}{2020}), \eprint{2003.07788}.

\bibitem[{\citenamefont{Kumar and Ghosh}(2020{\natexlab{a}})}]{Kumar:2020owy}
\bibinfo{author}{\bibfnamefont{R.}~\bibnamefont{Kumar}} \bibnamefont{and}
  \bibinfo{author}{\bibfnamefont{S.~G.} \bibnamefont{Ghosh}}
  (\bibinfo{year}{2020}{\natexlab{a}}), \eprint{2003.08927}.

\bibitem[{\citenamefont{Ghosh and Maharaj}(2020)}]{Ghosh:2020vpc}
\bibinfo{author}{\bibfnamefont{S.~G.} \bibnamefont{Ghosh}} \bibnamefont{and}
  \bibinfo{author}{\bibfnamefont{S.~D.} \bibnamefont{Maharaj}}
  (\bibinfo{year}{2020}), \eprint{2003.09841}.

\bibitem[{\citenamefont{Ghosh and Kumar}(2020)}]{Ghosh:2020syx}
\bibinfo{author}{\bibfnamefont{S.~G.} \bibnamefont{Ghosh}} \bibnamefont{and}
  \bibinfo{author}{\bibfnamefont{R.}~\bibnamefont{Kumar}}
  (\bibinfo{year}{2020}), \eprint{2003.12291}.

\bibitem[{\citenamefont{Kumar and Kumar}(2020)}]{Kumar:2020uyz}
\bibinfo{author}{\bibfnamefont{A.}~\bibnamefont{Kumar}} \bibnamefont{and}
  \bibinfo{author}{\bibfnamefont{R.}~\bibnamefont{Kumar}}
  (\bibinfo{year}{2020}), \eprint{2003.13104}.

\bibitem[{\citenamefont{Kumar and Ghosh}(2020{\natexlab{b}})}]{Kumar:2020xvu}
\bibinfo{author}{\bibfnamefont{A.}~\bibnamefont{Kumar}} \bibnamefont{and}
  \bibinfo{author}{\bibfnamefont{S.~G.} \bibnamefont{Ghosh}}
  (\bibinfo{year}{2020}{\natexlab{b}}), \eprint{2004.01131}.

\bibitem[{\citenamefont{Heydari-Fard et~al.}(2020)\citenamefont{Heydari-Fard,
  Heydari-Fard, and Sepangi}}]{Heydari-Fard:2020sib}
\bibinfo{author}{\bibfnamefont{M.}~\bibnamefont{Heydari-Fard}},
  \bibinfo{author}{\bibfnamefont{M.}~\bibnamefont{Heydari-Fard}},
  \bibnamefont{and} \bibinfo{author}{\bibfnamefont{H.}~\bibnamefont{Sepangi}}
  (\bibinfo{year}{2020}), \eprint{2004.02140}.

\bibitem[{\citenamefont{Hegde et~al.}(2020)\citenamefont{Hegde, Naveena~Kumara,
  Rizwan, M., and Ali}}]{Hegde:2020xlv}
\bibinfo{author}{\bibfnamefont{K.}~\bibnamefont{Hegde}},
  \bibinfo{author}{\bibfnamefont{A.}~\bibnamefont{Naveena~Kumara}},
  \bibinfo{author}{\bibfnamefont{C.~A.} \bibnamefont{Rizwan}},
  \bibinfo{author}{\bibfnamefont{A.~K.} \bibnamefont{M.}}, \bibnamefont{and}
  \bibinfo{author}{\bibfnamefont{M.~S.} \bibnamefont{Ali}}
  (\bibinfo{year}{2020}), \eprint{2003.08778}.

\bibitem[{\citenamefont{Hosseini~Mansoori}(2020)}]{HosseiniMansoori:2020yfj}
\bibinfo{author}{\bibfnamefont{S.~A.} \bibnamefont{Hosseini~Mansoori}}
  (\bibinfo{year}{2020}), \eprint{2003.13382}.

\bibitem[{\citenamefont{Wei and Liu}(2020)}]{Wei:2020poh}
\bibinfo{author}{\bibfnamefont{S.-W.} \bibnamefont{Wei}} \bibnamefont{and}
  \bibinfo{author}{\bibfnamefont{Y.-X.} \bibnamefont{Liu}}
  (\bibinfo{year}{2020}), \eprint{2003.14275}.

\bibitem[{\citenamefont{Eslam~Panah and Jafarzade}(2020)}]{EslamPanah:2020hoj}
\bibinfo{author}{\bibfnamefont{B.}~\bibnamefont{Eslam~Panah}} \bibnamefont{and}
  \bibinfo{author}{\bibfnamefont{K.}~\bibnamefont{Jafarzade}}
  (\bibinfo{year}{2020}), \eprint{2004.04058}.

\bibitem[{\citenamefont{Konoplya and Zinhailo}(2020)}]{Konoplya:2020bxa}
\bibinfo{author}{\bibfnamefont{R.}~\bibnamefont{Konoplya}} \bibnamefont{and}
  \bibinfo{author}{\bibfnamefont{A.}~\bibnamefont{Zinhailo}}
  (\bibinfo{year}{2020}), \eprint{2003.01188}.

\bibitem[{\citenamefont{Churilova}(2020)}]{Churilova:2020aca}
\bibinfo{author}{\bibfnamefont{M.}~\bibnamefont{Churilova}}
  (\bibinfo{year}{2020}), \eprint{2004.00513}.

\bibitem[{\citenamefont{Mishra}(2020)}]{Mishra:2020gce}
\bibinfo{author}{\bibfnamefont{A.~K.} \bibnamefont{Mishra}}
  (\bibinfo{year}{2020}), \eprint{2004.01243}.

\bibitem[{\citenamefont{Aragón et~al.}(2020)\citenamefont{Aragón, Bécar,
  González, and Vásquez}}]{Aragon:2020qdc}
\bibinfo{author}{\bibfnamefont{A.}~\bibnamefont{Aragón}},
  \bibinfo{author}{\bibfnamefont{R.}~\bibnamefont{Bécar}},
  \bibinfo{author}{\bibfnamefont{P.}~\bibnamefont{González}},
  \bibnamefont{and} \bibinfo{author}{\bibfnamefont{Y.}~\bibnamefont{Vásquez}}
  (\bibinfo{year}{2020}), \eprint{2004.05632}.

\bibitem[{\citenamefont{Doneva and Yazadjiev}(2020)}]{Doneva:2020ped}
\bibinfo{author}{\bibfnamefont{D.~D.} \bibnamefont{Doneva}} \bibnamefont{and}
  \bibinfo{author}{\bibfnamefont{S.~S.} \bibnamefont{Yazadjiev}}
  (\bibinfo{year}{2020}), \eprint{2003.10284}.

\bibitem[{\citenamefont{Banerjee and Singh}(2020)}]{Banerjee:2020stc}
\bibinfo{author}{\bibfnamefont{A.}~\bibnamefont{Banerjee}} \bibnamefont{and}
  \bibinfo{author}{\bibfnamefont{K.~N.} \bibnamefont{Singh}}
  (\bibinfo{year}{2020}), \eprint{2005.04028}.

\bibitem[{\citenamefont{Narain and Zhang}(2020)}]{1795133}
\bibinfo{author}{\bibfnamefont{G.}~\bibnamefont{Narain}} \bibnamefont{and}
  \bibinfo{author}{\bibfnamefont{H.-Q.} \bibnamefont{Zhang}}
  (\bibinfo{year}{2020}), \eprint{2005.05183}.

\bibitem[{\citenamefont{Odintsov et~al.}(2020)\citenamefont{Odintsov,
  Oikonomou, and Fronimos}}]{Odintsov:2020sqy}
\bibinfo{author}{\bibfnamefont{S.}~\bibnamefont{Odintsov}},
  \bibinfo{author}{\bibfnamefont{V.}~\bibnamefont{Oikonomou}},
  \bibnamefont{and} \bibinfo{author}{\bibfnamefont{F.}~\bibnamefont{Fronimos}}
  (\bibinfo{year}{2020}), \eprint{2003.13724}.

\bibitem[{\citenamefont{Odintsov and Oikonomou}(2020)}]{Odintsov:2020zkl}
\bibinfo{author}{\bibfnamefont{S.}~\bibnamefont{Odintsov}} \bibnamefont{and}
  \bibinfo{author}{\bibfnamefont{V.}~\bibnamefont{Oikonomou}},
  \bibinfo{journal}{Phys. Lett. B} \textbf{\bibinfo{volume}{805}},
  \bibinfo{pages}{135437} (\bibinfo{year}{2020}), \eprint{2004.00479}.

\bibitem[{\citenamefont{Jusufi et~al.}(2020)\citenamefont{Jusufi, Banerjee, and
  Ghosh}}]{Jusufi:2020yus}
\bibinfo{author}{\bibfnamefont{K.}~\bibnamefont{Jusufi}},
  \bibinfo{author}{\bibfnamefont{A.}~\bibnamefont{Banerjee}}, \bibnamefont{and}
  \bibinfo{author}{\bibfnamefont{S.~G.} \bibnamefont{Ghosh}}
  (\bibinfo{year}{2020}), \eprint{2004.10750}.

\bibitem[{\citenamefont{Liu et~al.}(2020)\citenamefont{Liu, Niu, Wang, and
  Zhang}}]{Liu:2020yhu}
\bibinfo{author}{\bibfnamefont{P.}~\bibnamefont{Liu}},
  \bibinfo{author}{\bibfnamefont{C.}~\bibnamefont{Niu}},
  \bibinfo{author}{\bibfnamefont{X.}~\bibnamefont{Wang}}, \bibnamefont{and}
  \bibinfo{author}{\bibfnamefont{C.-Y.} \bibnamefont{Zhang}}
  (\bibinfo{year}{2020}), \eprint{2004.14267}.

\bibitem[{\citenamefont{Aoki et~al.}(2020)\citenamefont{Aoki, Gorji, and
  Mukohyama}}]{Aoki:2020lig}
\bibinfo{author}{\bibfnamefont{K.}~\bibnamefont{Aoki}},
  \bibinfo{author}{\bibfnamefont{M.~A.} \bibnamefont{Gorji}}, \bibnamefont{and}
  \bibinfo{author}{\bibfnamefont{S.}~\bibnamefont{Mukohyama}}
  (\bibinfo{year}{2020}), \eprint{2005.03859}.

\bibitem[{\citenamefont{Arrechea et~al.}(2020)\citenamefont{Arrechea, Delhom,
  and Jiménez-Cano}}]{Arrechea:2020evj}
\bibinfo{author}{\bibfnamefont{J.}~\bibnamefont{Arrechea}},
  \bibinfo{author}{\bibfnamefont{A.}~\bibnamefont{Delhom}}, \bibnamefont{and}
  \bibinfo{author}{\bibfnamefont{A.}~\bibnamefont{Jiménez-Cano}}
  (\bibinfo{year}{2020}), \eprint{2004.12998}.

\bibitem[{\citenamefont{Lu and Pang}(2020)}]{Lu:2020iav}
\bibinfo{author}{\bibfnamefont{H.}~\bibnamefont{Lu}} \bibnamefont{and}
  \bibinfo{author}{\bibfnamefont{Y.}~\bibnamefont{Pang}}
  (\bibinfo{year}{2020}), \eprint{2003.11552}.

\bibitem[{\citenamefont{Gurses et~al.}(2020)\citenamefont{Gurses, Sisman, and
  Tekin}}]{Gurses:2020ofy}
\bibinfo{author}{\bibfnamefont{M.}~\bibnamefont{Gurses}},
  \bibinfo{author}{\bibfnamefont{T.~C.} \bibnamefont{Sisman}},
  \bibnamefont{and} \bibinfo{author}{\bibfnamefont{B.}~\bibnamefont{Tekin}}
  (\bibinfo{year}{2020}), \eprint{2004.03390}.

\bibitem[{\citenamefont{Ai}(2020)}]{Ai:2020peo}
\bibinfo{author}{\bibfnamefont{W.-Y.} \bibnamefont{Ai}} (\bibinfo{year}{2020}),
  \eprint{2004.02858}.

\bibitem[{\citenamefont{Mahapatra}(2020)}]{Mahapatra:2020rds}
\bibinfo{author}{\bibfnamefont{S.}~\bibnamefont{Mahapatra}}
  (\bibinfo{year}{2020}), \eprint{2004.09214}.

\bibitem[{\citenamefont{Van~Acoleyen and
  Van~Doorsselaere}(2011)}]{VanAcoleyen11-1}
\bibinfo{author}{\bibfnamefont{K.}~\bibnamefont{Van~Acoleyen}}
  \bibnamefont{and}
  \bibinfo{author}{\bibfnamefont{J.}~\bibnamefont{Van~Doorsselaere}},
  \bibinfo{journal}{Phys. Rev.} \textbf{\bibinfo{volume}{D83}},
  \bibinfo{pages}{084025} (\bibinfo{year}{2011}), \eprint{1102.0487}.

\bibitem[{\citenamefont{Charmousis}(2015)}]{Charmousis:2014mia}
\bibinfo{author}{\bibfnamefont{C.}~\bibnamefont{Charmousis}},
  \bibinfo{journal}{Lect.\ Notes Phys.} \textbf{\bibinfo{volume}{892}},
  \bibinfo{pages}{25} (\bibinfo{year}{2015}), \eprint{1405.1612}.

\bibitem[{\citenamefont{Horndeski}(1974)}]{Horndeski:1974wa}
\bibinfo{author}{\bibfnamefont{G.~W.} \bibnamefont{Horndeski}},
  \bibinfo{journal}{Int. J. Theor. Phys.} \textbf{\bibinfo{volume}{10}},
  \bibinfo{pages}{363} (\bibinfo{year}{1974}).

\bibitem[{\citenamefont{Fernandes et~al.}(2020)\citenamefont{Fernandes,
  Carrilho, Clifton, and Mulryne}}]{Fernandes:2020nbq}
\bibinfo{author}{\bibfnamefont{P.~G.} \bibnamefont{Fernandes}},
  \bibinfo{author}{\bibfnamefont{P.}~\bibnamefont{Carrilho}},
  \bibinfo{author}{\bibfnamefont{T.}~\bibnamefont{Clifton}}, \bibnamefont{and}
  \bibinfo{author}{\bibfnamefont{D.~J.} \bibnamefont{Mulryne}}
  (\bibinfo{year}{2020}), \eprint{2004.08362}.

\bibitem[{\citenamefont{Hennigar
  et~al.}(2020{\natexlab{a}})\citenamefont{Hennigar, Kubiznak, Mann, and
  Pollack}}]{Hennigar:2020lsl}
\bibinfo{author}{\bibfnamefont{R.~A.} \bibnamefont{Hennigar}},
  \bibinfo{author}{\bibfnamefont{D.}~\bibnamefont{Kubiznak}},
  \bibinfo{author}{\bibfnamefont{R.~B.} \bibnamefont{Mann}}, \bibnamefont{and}
  \bibinfo{author}{\bibfnamefont{C.}~\bibnamefont{Pollack}}
  (\bibinfo{year}{2020}{\natexlab{a}}), \eprint{2004.09472}.

\bibitem[{\citenamefont{Mann and Ross}(1993)}]{Mann:1992ar}
\bibinfo{author}{\bibfnamefont{R.~B.} \bibnamefont{Mann}} \bibnamefont{and}
  \bibinfo{author}{\bibfnamefont{S.}~\bibnamefont{Ross}},
  \bibinfo{journal}{Class.\ Quant.\ Grav.} \textbf{\bibinfo{volume}{10}},
  \bibinfo{pages}{1405} (\bibinfo{year}{1993}), \eprint{gr-qc/9208004}.

\bibitem[{\citenamefont{Zwiebach}(1985)}]{Zwiebach:1985uq}
\bibinfo{author}{\bibfnamefont{B.}~\bibnamefont{Zwiebach}},
  \bibinfo{journal}{Phys. Lett. B} \textbf{\bibinfo{volume}{156}},
  \bibinfo{pages}{315} (\bibinfo{year}{1985}).

\bibitem[{\citenamefont{Duff et~al.}(1986)\citenamefont{Duff, Nilsson, and
  Pope}}]{Duff86-1}
\bibinfo{author}{\bibfnamefont{M.~J.} \bibnamefont{Duff}},
  \bibinfo{author}{\bibfnamefont{B.~E.~W.} \bibnamefont{Nilsson}},
  \bibnamefont{and} \bibinfo{author}{\bibfnamefont{C.~N.} \bibnamefont{Pope}},
  \bibinfo{journal}{Phys. Rept.} \textbf{\bibinfo{volume}{130}},
  \bibinfo{pages}{1} (\bibinfo{year}{1986}).

\bibitem[{\citenamefont{Metsaev and Tseytlin}(1987)}]{Metsaev87-1}
\bibinfo{author}{\bibfnamefont{R.~R.} \bibnamefont{Metsaev}} \bibnamefont{and}
  \bibinfo{author}{\bibfnamefont{A.~A.} \bibnamefont{Tseytlin}},
  \bibinfo{journal}{Nucl. Phys.} \textbf{\bibinfo{volume}{B293}},
  \bibinfo{pages}{385} (\bibinfo{year}{1987}).

\bibitem[{\citenamefont{Grumiller and Jackiw}(2007)}]{Grumiller:2007wb}
\bibinfo{author}{\bibfnamefont{D.}~\bibnamefont{Grumiller}} \bibnamefont{and}
  \bibinfo{author}{\bibfnamefont{R.}~\bibnamefont{Jackiw}}
  (\bibinfo{year}{2007}), \eprint{0712.3775}.

\bibitem[{\citenamefont{Frassino et~al.}(2015)\citenamefont{Frassino, Mann, and
  Mureika}}]{Frassino:2015oca}
\bibinfo{author}{\bibfnamefont{A.~M.} \bibnamefont{Frassino}},
  \bibinfo{author}{\bibfnamefont{R.~B.} \bibnamefont{Mann}}, \bibnamefont{and}
  \bibinfo{author}{\bibfnamefont{J.~R.} \bibnamefont{Mureika}},
  \bibinfo{journal}{Phys. Rev. D} \textbf{\bibinfo{volume}{92}},
  \bibinfo{pages}{124069} (\bibinfo{year}{2015}), \eprint{1509.05481}.

\bibitem[{\citenamefont{Rosso and Svesko}(2020)}]{Rosso:2020zkk}
\bibinfo{author}{\bibfnamefont{F.}~\bibnamefont{Rosso}} \bibnamefont{and}
  \bibinfo{author}{\bibfnamefont{A.}~\bibnamefont{Svesko}}
  (\bibinfo{year}{2020}), \eprint{2003.10462}.

\bibitem[{\citenamefont{Dabrowski et~al.}(2009)\citenamefont{Dabrowski,
  Garecki, and Blaschke}}]{Dabrowski:2008kx}
\bibinfo{author}{\bibfnamefont{M.~P.} \bibnamefont{Dabrowski}},
  \bibinfo{author}{\bibfnamefont{J.}~\bibnamefont{Garecki}}, \bibnamefont{and}
  \bibinfo{author}{\bibfnamefont{D.~B.} \bibnamefont{Blaschke}},
  \bibinfo{journal}{Annalen Phys.} \textbf{\bibinfo{volume}{18}},
  \bibinfo{pages}{13} (\bibinfo{year}{2009}), \eprint{0806.2683}.

\bibitem[{\citenamefont{Hennigar
  et~al.}(2020{\natexlab{b}})\citenamefont{Hennigar, Kubiznak, Mann, and
  Pollack}}]{Hennigar:2020fkv}
\bibinfo{author}{\bibfnamefont{R.~A.} \bibnamefont{Hennigar}},
  \bibinfo{author}{\bibfnamefont{D.}~\bibnamefont{Kubiznak}},
  \bibinfo{author}{\bibfnamefont{R.~B.} \bibnamefont{Mann}}, \bibnamefont{and}
  \bibinfo{author}{\bibfnamefont{C.}~\bibnamefont{Pollack}}
  (\bibinfo{year}{2020}{\natexlab{b}}), \eprint{2004.12995}.

\bibitem[{\citenamefont{Deffayet et~al.}(2009)\citenamefont{Deffayet, Deser,
  and Esposito-Farese}}]{Deffayet:2009mn}
\bibinfo{author}{\bibfnamefont{C.}~\bibnamefont{Deffayet}},
  \bibinfo{author}{\bibfnamefont{S.}~\bibnamefont{Deser}}, \bibnamefont{and}
  \bibinfo{author}{\bibfnamefont{G.}~\bibnamefont{Esposito-Farese}},
  \bibinfo{journal}{Phys. Rev. D} \textbf{\bibinfo{volume}{80}},
  \bibinfo{pages}{064015} (\bibinfo{year}{2009}), \eprint{0906.1967}.

\bibitem[{\citenamefont{Charmousis et~al.}(2012)\citenamefont{Charmousis,
  Gouteraux, and Kiritsis}}]{Charmousis12-1}
\bibinfo{author}{\bibfnamefont{C.}~\bibnamefont{Charmousis}},
  \bibinfo{author}{\bibfnamefont{B.}~\bibnamefont{Gouteraux}},
  \bibnamefont{and} \bibinfo{author}{\bibfnamefont{E.}~\bibnamefont{Kiritsis}},
  \bibinfo{journal}{JHEP} \textbf{\bibinfo{volume}{09}}, \bibinfo{pages}{011}
  (\bibinfo{year}{2012}), \eprint{1206.1499}.

\bibitem[{\citenamefont{Green et~al.}(2012)\citenamefont{Green, Schwarz, and
  Witten}}]{Green12-1}
\bibinfo{author}{\bibfnamefont{M.~B.} \bibnamefont{Green}},
  \bibinfo{author}{\bibfnamefont{J.~H.} \bibnamefont{Schwarz}},
  \bibnamefont{and} \bibinfo{author}{\bibfnamefont{E.}~\bibnamefont{Witten}},
  \emph{\bibinfo{title}{{Superstring Theory Vol. 1}}}
  (\bibinfo{publisher}{Cambridge University Press}, \bibinfo{year}{2012}).

\bibitem[{\citenamefont{Gasperini}(2007)}]{Gasperini07-1}
\bibinfo{author}{\bibfnamefont{M.}~\bibnamefont{Gasperini}},
  \emph{\bibinfo{title}{{Elements of string cosmology}}}
  (\bibinfo{publisher}{Cambridge University Press}, \bibinfo{year}{2007}).

\bibitem[{\citenamefont{Manton et~al.}({2020})\citenamefont{Manton, Parikh, and
  Svesko}}]{TMA20}
\bibinfo{author}{\bibfnamefont{T.}~\bibnamefont{Manton}},
  \bibinfo{author}{\bibfnamefont{M.}~\bibnamefont{Parikh}}, \bibnamefont{and}
  \bibinfo{author}{\bibfnamefont{A.}~\bibnamefont{Svesko}},
  \emph{\bibinfo{title}{{Unpublished}}} (\bibinfo{year}{{2020}}).

\bibitem[{\citenamefont{Padilla and Sivanesan}(2013)}]{Padilla:2012dx}
\bibinfo{author}{\bibfnamefont{A.}~\bibnamefont{Padilla}} \bibnamefont{and}
  \bibinfo{author}{\bibfnamefont{V.}~\bibnamefont{Sivanesan}},
  \bibinfo{journal}{JHEP} \textbf{\bibinfo{volume}{04}}, \bibinfo{pages}{032}
  (\bibinfo{year}{2013}), \eprint{1210.4026}.

\bibitem[{\citenamefont{Kobayashi et~al.}(2013)\citenamefont{Kobayashi,
  Tanahashi, and Yamaguchi}}]{Kobayashi:2013ina}
\bibinfo{author}{\bibfnamefont{T.}~\bibnamefont{Kobayashi}},
  \bibinfo{author}{\bibfnamefont{N.}~\bibnamefont{Tanahashi}},
  \bibnamefont{and}
  \bibinfo{author}{\bibfnamefont{M.}~\bibnamefont{Yamaguchi}},
  \bibinfo{journal}{Phys. Rev. D} \textbf{\bibinfo{volume}{88}},
  \bibinfo{pages}{083504} (\bibinfo{year}{2013}), \eprint{1308.4798}.

\bibitem[{\citenamefont{Easson et~al.}(2013)\citenamefont{Easson, Sawicki, and
  Vikman}}]{Easson:2013bda}
\bibinfo{author}{\bibfnamefont{D.~A.} \bibnamefont{Easson}},
  \bibinfo{author}{\bibfnamefont{I.}~\bibnamefont{Sawicki}}, \bibnamefont{and}
  \bibinfo{author}{\bibfnamefont{A.}~\bibnamefont{Vikman}},
  \bibinfo{journal}{JCAP} \textbf{\bibinfo{volume}{07}}, \bibinfo{pages}{014}
  (\bibinfo{year}{2013}), \eprint{1304.3903}.

\bibitem[{\citenamefont{Creminelli et~al.}(2010)\citenamefont{Creminelli,
  Nicolis, and Trincherini}}]{Creminelli:2010ba}
\bibinfo{author}{\bibfnamefont{P.}~\bibnamefont{Creminelli}},
  \bibinfo{author}{\bibfnamefont{A.}~\bibnamefont{Nicolis}}, \bibnamefont{and}
  \bibinfo{author}{\bibfnamefont{E.}~\bibnamefont{Trincherini}},
  \bibinfo{journal}{JCAP} \textbf{\bibinfo{volume}{11}}, \bibinfo{pages}{021}
  (\bibinfo{year}{2010}), \eprint{1007.0027}.

\bibitem[{\citenamefont{Kobayashi et~al.}(2010)\citenamefont{Kobayashi,
  Yamaguchi, and Yokoyama}}]{Kobayashi:2010cm}
\bibinfo{author}{\bibfnamefont{T.}~\bibnamefont{Kobayashi}},
  \bibinfo{author}{\bibfnamefont{M.}~\bibnamefont{Yamaguchi}},
  \bibnamefont{and} \bibinfo{author}{\bibfnamefont{J.}~\bibnamefont{Yokoyama}},
  \bibinfo{journal}{Phys. Rev. Lett.} \textbf{\bibinfo{volume}{105}},
  \bibinfo{pages}{231302} (\bibinfo{year}{2010}), \eprint{1008.0603}.

\bibitem[{\citenamefont{Qiu et~al.}(2011)\citenamefont{Qiu, Evslin, Cai, Li,
  and Zhang}}]{Qiu:2011cy}
\bibinfo{author}{\bibfnamefont{T.}~\bibnamefont{Qiu}},
  \bibinfo{author}{\bibfnamefont{J.}~\bibnamefont{Evslin}},
  \bibinfo{author}{\bibfnamefont{Y.-F.} \bibnamefont{Cai}},
  \bibinfo{author}{\bibfnamefont{M.}~\bibnamefont{Li}}, \bibnamefont{and}
  \bibinfo{author}{\bibfnamefont{X.}~\bibnamefont{Zhang}},
  \bibinfo{journal}{JCAP} \textbf{\bibinfo{volume}{10}}, \bibinfo{pages}{036}
  (\bibinfo{year}{2011}), \eprint{1108.0593}.

\bibitem[{\citenamefont{Easson et~al.}(2011)\citenamefont{Easson, Sawicki, and
  Vikman}}]{Easson:2011zy}
\bibinfo{author}{\bibfnamefont{D.~A.} \bibnamefont{Easson}},
  \bibinfo{author}{\bibfnamefont{I.}~\bibnamefont{Sawicki}}, \bibnamefont{and}
  \bibinfo{author}{\bibfnamefont{A.}~\bibnamefont{Vikman}},
  \bibinfo{journal}{JCAP} \textbf{\bibinfo{volume}{11}}, \bibinfo{pages}{021}
  (\bibinfo{year}{2011}), \eprint{1109.1047}.

\bibitem[{\citenamefont{Deffayet et~al.}(2010)\citenamefont{Deffayet, Pujolas,
  Sawicki, and Vikman}}]{Deffayet:2010qz}
\bibinfo{author}{\bibfnamefont{C.}~\bibnamefont{Deffayet}},
  \bibinfo{author}{\bibfnamefont{O.}~\bibnamefont{Pujolas}},
  \bibinfo{author}{\bibfnamefont{I.}~\bibnamefont{Sawicki}}, \bibnamefont{and}
  \bibinfo{author}{\bibfnamefont{A.}~\bibnamefont{Vikman}},
  \bibinfo{journal}{JCAP} \textbf{\bibinfo{volume}{10}}, \bibinfo{pages}{026}
  (\bibinfo{year}{2010}), \eprint{1008.0048}.

\bibitem[{\citenamefont{Elder et~al.}(2014)\citenamefont{Elder, Joyce, and
  Khoury}}]{Elder:2013gya}
\bibinfo{author}{\bibfnamefont{B.}~\bibnamefont{Elder}},
  \bibinfo{author}{\bibfnamefont{A.}~\bibnamefont{Joyce}}, \bibnamefont{and}
  \bibinfo{author}{\bibfnamefont{J.}~\bibnamefont{Khoury}},
  \bibinfo{journal}{Phys. Rev. D} \textbf{\bibinfo{volume}{89}},
  \bibinfo{pages}{044027} (\bibinfo{year}{2014}), \eprint{1311.5889}.

\bibitem[{\citenamefont{Rubakov}(2014)}]{Rubakov:2014jja}
\bibinfo{author}{\bibfnamefont{V.}~\bibnamefont{Rubakov}},
  \bibinfo{journal}{Usp. Fiz. Nauk} \textbf{\bibinfo{volume}{184}},
  \bibinfo{pages}{137} (\bibinfo{year}{2014}), \eprint{1401.4024}.

\bibitem[{\citenamefont{Rubakov}(2016{\natexlab{a}})}]{Rubakov:2015gza}
\bibinfo{author}{\bibfnamefont{V.}~\bibnamefont{Rubakov}},
  \bibinfo{journal}{Teor. Mat. Fiz.} \textbf{\bibinfo{volume}{187}},
  \bibinfo{pages}{338} (\bibinfo{year}{2016}{\natexlab{a}}),
  \eprint{1509.08808}.

\bibitem[{\citenamefont{Rubakov}(2016{\natexlab{b}})}]{Rubakov:2016zah}
\bibinfo{author}{\bibfnamefont{V.}~\bibnamefont{Rubakov}},
  \bibinfo{journal}{Theor. Math. Phys.} \textbf{\bibinfo{volume}{188}},
  \bibinfo{pages}{1253} (\bibinfo{year}{2016}{\natexlab{b}}),
  \eprint{1601.06566}.

\bibitem[{\citenamefont{Gannouji and Sami}(2012)}]{Gannouji:2011qz}
\bibinfo{author}{\bibfnamefont{R.}~\bibnamefont{Gannouji}} \bibnamefont{and}
  \bibinfo{author}{\bibfnamefont{M.}~\bibnamefont{Sami}},
  \bibinfo{journal}{Phys. Rev. D} \textbf{\bibinfo{volume}{85}},
  \bibinfo{pages}{024019} (\bibinfo{year}{2012}), \eprint{1107.1892}.

\end{thebibliography}

\end{document}